\documentclass[aps,pra,english,twocolumn,showpacs,preprintnumbers,amsmath,amssymb,floatfix,superscriptaddress,longbibliography,nofootinbib]{revtex4-1}

\usepackage[T1]{fontenc}
\usepackage[latin9]{inputenc}
\usepackage{graphicx}
\usepackage{amssymb}
\usepackage{babel}
\makeatletter

\usepackage[version=3]{mhchem}

\usepackage{color}

\usepackage{setspace}

\global\arraycolsep=2pt
\usepackage{stmaryrd}
\usepackage{amsmath}
\usepackage{amssymb}
\usepackage{graphicx}
\usepackage{textcomp}
\usepackage{calrsfs}
\usepackage{yfonts}
\usepackage{bm,braket}
\usepackage{color}

\newcommand{\ben}{\begin{equation*}}
\newcommand{\een}{\end{equation*}}
\newcommand{\bean}{\begin{eqnarray*}}
\newcommand{\eean}{\end{eqnarray*}}

\newcommand{\be}{\begin{equation}}
\newcommand{\ee}{\end{equation}}
\newcommand{\bea}{\begin{eqnarray}}
\newcommand{\eea}{\end{eqnarray}}

\DeclareMathOperator{\Tr}{Tr}
\newcommand{\tr}{\text{tr}}

\newcommand{\mRe}{\text{Re}}

\newcommand{\diag}{\text{diag}}

\newcommand{\psumbar}{\sum\nolimits^\prime}
\DeclareMathOperator*{\psum}{\psumbar}

\makeatother
\usepackage{babel}
\makeatother

\begin{document}
\title{Thermal Casimir interactions in multi-particle systems: scattering channel approach}

\author{Yang Li}
  \email{leon@ncu.edu.cn}
  \affiliation{School of Physics and Materials Science, Nanchang University, Nanchang 330031, China}
  \affiliation{Institute of Space Science and Technology, Nanchang University, Nanchang 330031, China}

\author{Kimball A. Milton}
  \email{kmilton@ou.edu}
  \affiliation{Homer L. Dodge Department of Physics and Astronomy, University of Oklahoma, Norman, Oklahoma 73019, USA}

\author{Iver Brevik}
  \email{iver.h.brevik@gmail.com}
  \affiliation{Department of Energy and Process Engineering, Norwegian University of Science and Technology, NO-7491 Trondheim, Norway}

\begin{abstract}
Multi-particle thermal Casimir interactions are investigated, mostly in terms of the Casimir entropy, from the point of view based on  multiple-scattering processes. The geometry of the scattering path is depicted in detail, and the contributions from different types of channels, namely the transverse, longitudinal and mixing channels, are demonstrated. The geometry of the path can strongly influence the weight of each channel in the path. Negativity and nonmonotonicity are commonly seen in the multi-particle Casimir entropy, the sources of which are diverse, including the geometry of the path, the types of polarization mixing, the polarizability of each particle, etc. Thermal contributions from
multi-particle scatterings can be significant in the system, while the zero-temperature multi-particle scattering effects are insignificant. Limiting behaviors from a multi-particle configuration to a continuum are briefly explored.
\end{abstract}

\date{\today}

\maketitle

\section{Introduction}
\label{I}
\par Since Casimir~\cite{casimir1948attraction} proposed the existence of a measurable attractive force, arising from the change of the zero-point energy due to two neutral perfectly conducting plates, in 1948, those phenomena, caused by the fluctuation of quantum fields and generally known as  Casimir effects~\cite{milton2003casimir,bordag2009advances,Dalvit2011Casimir}, have been studied and probed both theoretically and experimentally. The applications of Casimir interactions, for instance in
nano-mechanics~\cite{andrews2014role,rodriguez2015classical}, are crucial. In  Casimir interactions, the multi-body contributions are important in many practical scenarios. To name only a few, recently the premelting of ice~\cite{esteso2020premelting,luengo2021lifshitz} and the levitation of a
nanoplate~\cite{zhao2019stable}, resulting from the Casimir interactions in multi-layer configurations, have been investigated, highlighting the nonmonotonicity of the Casimir stress or force. Indeed, the nonmonotonicity of Casimir interactions is commonly seen for many-body systems and has been attracting more and more attention~\cite{Schaden2011Irreducible,Shajesh2011Many,rodriguez2009three,shajesh2012significance,Milton2015Casimir}, which promises theoretical advances and applications. Remarkably, an experimental study on the multi-body Casimir interaction in the sphere-plate-sphere is carried out very recently~\cite{xu2022observation}, which may lead to the ability to control the multi-body Casimir interactions for practical applications. In another exciting direction, the importance of the multi-body Casimir interaction in mesoscopic systems has recently been recognized, and investigated in the context of low-dimensional carbon allotropes close to a gold surface~\cite{venkataram2019impact}.


\par Although there have been  many theoretical and practical advances, there are still aspects of the multi-body Casimir interaction incompletely examined, among which the multi-body thermal effects, or the influences of thermal fluctuations, are very important. At present, the importance of thermal Casimir effects is widely recognized, and
research on the topic of  Casimir entropy, which is an explicit manifestation of thermal
Casimir effects, has been attracting more and more attention. In fact, Casimir entropy,
the entropy of the quantum field induced by the presence of boundaries or media, has been under investigation for many years. 
For example, the Casimir entropy is invoked in the continuing controversy concerning the proper description of the permittivity of a metal when evaluating the Casimir force between metal plates at finite temperature, studied both experimentally~\cite{Decca2005Precise,Liu2019Examining,Sushkov2011Observation,Garcia2012Casimir} and theoretically~\cite{bezerra2004violation,klimchitskaya2017low,brevik2006thermal,milton2012thermal}. 
Some theoretical works claimed that the Drude model violates the third law of thermodynamics~\cite{bezerra2004violation,klimchitskaya2017low} in that the Casimir entropy does not go to zero as the temperature vanishes, while others argued that the Drude model does not cause any such difficulty~\cite{brevik2006thermal,milton2012thermal}. For a review on this continuing controversy,  the reader is referred to Ref.~[\onlinecite{Dalvit2011Casimir}]. More recently, researchers,
focusing on Casimir entropy, discovered interesting properties, featuring its negativity. The Casimir entropy due to the interaction between two bodies can be negative, which was thought to be induced by the dissipation of the media~\cite{bezerra2002thermodynamical,bezerra2002correlation}, as well as by the geometry
alone~\cite{canaguier2010thermal,rodriguez2011casimir,ingold2015geometric,milton2015negative,milton2017negative}. For a single body, its Casimir self-entropy~\cite{li2016casimir,milton2017casimir,bordag2018free,bordag2018entropy} is also not trivial, and can have surprising properties, such as negativity, nonmonotonicity, or even divergences, to be dealt with properly. All those findings suggest more physics in  thermal Casimir interactions is yet to be unveiled.

\par In previous investigations of the multi-body Casimir interaction~\cite{rodriguez2009three,shajesh2012significance,milton2013three,milton2015three}, attention was mostly paid to the zero-temperature Casimir interaction,
where nonmonotonicity of the Casimir force was found.
Notably, some specific scattering channels, in some systems with relatively simple geometric configurations, were picked out there as objects of study, and compared with 
pairwise results {as reference. On the other hand, the scattering approach has been used to analyze thermal Casimir effects, including the Casimir entropy. For instance, in Ref.~[\onlinecite{ingold2015geometric}], Ingold et al.\ studied the purely geometric origin of the negativity of Casimir entropy in the plane-sphere and sphere-sphere configurations, and the role of polarization-mixing channels in the appearance of this negativity is emphasized, which we confirm here as well. In an effort to enhance understanding of thermal Casimir interactions in multi-body systems, in this paper, we 
take the perspective of multi-scattering processes, and explore  thermal Casimir effects, particularly Casimir entropy, 
examining some theoretical models of interacting polarizable nanoparticles. We show that polarization-conserving channels, involving two or more particles, can also make negative contributions to the Casimir entropy. We describe the geometry of a general scattering path, and the corresponding influence on the thermal Casimir interaction. 
The multi-particle thermal contributions are, in general, non-negligible. Nonmonotonicity and negativity are almost unavoidable, implying the significance of multi-particle thermal Casimir interactions. Our scheme may provide an alternative approach to evaluating the thermal Casimir interactions in systems of complex geometry.

\par This paper is organized as follows: In Sec.~\ref{T}, the general theory employed is presented explicitly, based on which we evaluate and analyze the thermal Casimir effects in the two-particle, three-particle, and particle-lattice systems in Sec.~\ref{RD}. There, we rederive some results in consistency with those in previous works. The limiting behaviors from the nanoparticle system to the continuum are also discussed briefly, in agreement with well-known results. Concluding remarks are given in Sec.~\ref{C}.

%

\par The natural units $\hbar=\varepsilon_0=\mu_0=c=k_B=1$ are adopted throughout, unless specified otherwise.


\section{Theory}
\label{T}

\par It is well-known that the Casimir free energy of the electromagnetic field at a finite equilibrium temperature $T$, induced by a medium with the permittivity $\bm{\varepsilon}=\bm{1}+\mathbf{V}$ and permeability $\bm{\mu}=\bm{1}$, can be written symbolically as~\cite{li2016casimir,milton2017casimir}
\begin{eqnarray}
\label{eqT.1}
F
&=&
\frac{T}{2}\sum_{n=-\infty}^{\infty}\Tr\ln(\bm{1}+\bm{\Gamma}_{0}\mathbf{V}),
\end{eqnarray}
where the sum is over the indices of the Matsubara frequencies $\zeta_n=2\pi Tn$, and the vacuum Green's dyadic at the imaginary frequency $\zeta$, denoted as $\bm{\Gamma}_{0;\zeta}$, satisfies\footnote{In this paper, the sign of the Green's dyadic is reversed from that in Refs.~\onlinecite{li2016casimir,milton2017casimir}.}
\begin{eqnarray}
\label{eqT.2}
\bigg(
\bm{1}+\frac{\nabla\times\nabla\times\bm{1}}{\zeta^2}
\bigg)\cdot\bm{\Gamma}_{0;\zeta}(\mathbf{r},\mathbf{r}')
=
\bm{1}\delta(\mathbf{r}-\mathbf{r}').
\end{eqnarray}
In principle, Eq.~\eqref{eqT.1} applies to a general multi-particle, or even continuum, scenario. We mainly concentrate on  multi-particle systems here. When only one particle is in the vacuum, then $F$ is the Casimir self-free energy of this particle, which has been specifically investigated
recently~\cite{li2022casimir,li2021negativity,milton2017remarks,milton2017casimir}. For a
two-particle cases $\rm P_1$ and $\rm P_2$ in Fig.~\ref{figsystem} as an example,
$\mathbf{V}=\mathbf{V}_1+\mathbf{V}_2$ is just the sum of the susceptibilities of these two particles.
\begin{figure}[h]
  \centering
  \includegraphics[scale=1.2]{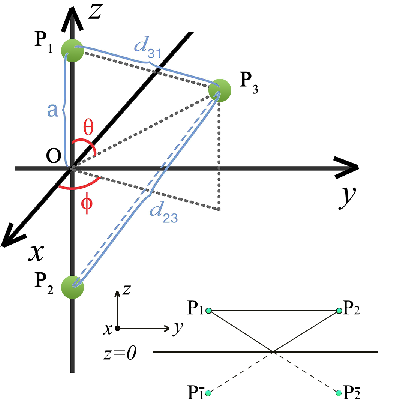}
  \caption{\label{figsystem}The schematic illustration of the three-particle configuration, which consists of three neutral polarizable particles. Particle 1 ($\rm P_1$) and Particle 2 ($\rm P_2$) are located at $\mathbf{R}_1=a\hat{\mathbf{z}}$ and $\mathbf{R}_2=-a\hat{\mathbf{z}}$, respectively. The distance between particles ${\rm P}_i,{\rm P}_j,\ i,j=1,2,3$ are denoted as $d_{ij}=d_{ji}$. Inset: Two atoms $\rm{P}_1$ and $\rm{P}_2$ in front of the perfectly conducting mirror located in the $z=0$ plane, and their images $\rm{P}_{\overline{1}}$ and $\rm{P}_{\overline{2}}$.}
\end{figure}
The corresponding Casimir free energy $F$ now is comprised of three parts, namely the Casimir self-free energy $F_i,\ i=1,2$ of $\rm P_1$ and $\rm P_2$, and the interacting contribution $F_{12}=F_{21}$, written as follows
\begin{equation}
\label{eqT.3}
F_{12}
=
F-F_1-F_2
=
\frac{T}{2}\sum_{n=-\infty}^{\infty}\Tr\ln
(
\bm{1}
-
\mathbf{t}_{1}
\mathbf{t}_{2}
)
,
\end{equation}
in which the scattering matrix $\mathbf{t}_i$ for ${\rm P}_i$ is defined as $\mathbf{t}_{i}=-\bm{\Gamma}_{0}\bm{\chi}_{i}$, and $\bm{\chi}_{i}=\mathbf{V}_{i}(\bm{1}+\bm{\Gamma}_{0}\mathbf{V}_{i})^{-1}$. It should be noted that the susceptibility of a particle in the vacuum would be modified by the self-interaction due to the fluctuating field. The parameters $\bm{\chi}_{i}$ and $\mathbf{V}_{i;\zeta}$ are regarded as the effective (or renormalized) and bare susceptibility of the particle~\cite{guo2021energetics}, respectively. The interaction free energy in Eq.~\eqref{eqT.3} involves the properties of both these particles, and has a quite simple form. For the three-particle case, like that schematized by Fig.~\ref{figsystem}, similar arguments apply, that is, the interacting Casimir free energy among three bodies can be expressed in terms of one-body and two-body free energies as $F_{123}=F-F^{(12)}-F^{(13)}-F_{23}+F_1$, but its explicit form is very complicated, as Ref.~[\onlinecite{milton2015three}] demonstrated. In our notation, $F_{123}$ is written as
\begin{eqnarray}
\label{eqT.4}
F_{123}
&=&
T\psum_{n=0}^{\infty}\Tr\ln
\bigg\{
\bm{1}
-
{\mathbf{x}}_{23}
\bigg[
(\bm{1}+\mathbf{t}_1)
\mathbf{x}_{21}
\mathbf{t}_{2}
(\bm{1}+\mathbf{t}_1)
\nonumber\\
& &
\times
\mathbf{x}_{31}
\mathbf{t}_{3}
-
\mathbf{t}_{2}\mathbf{t}_{3}
\bigg]
\bigg\}
,
\end{eqnarray}
with $\mathbf{x}_{ij;\zeta_n}=
(\bm{1}-\mathbf{t}_{i;\zeta_n}\mathbf{t}_{j;\zeta_n})^{-1}$. Employing formulas in Eqs.~\eqref{eqT.3} and \eqref{eqT.4}, one can extract the properties of thermal Casimir effects in two- and three-particle configurations. Particularly important are the Casimir entropy $S$, and the Casimir force $\mathcal{F}$ acting on the particle located at $\mathbf{r}$, for instance, which can be derived based on their relations with the corresponding Casimir free energy $F$
\begin{eqnarray}
\label{eqT.5}
S=-\frac{\partial F}{\partial T}
,\
\mathcal{F}=-\frac{\partial F}{\partial\mathbf{r}},
\end{eqnarray}
where $T$ is the temperature. To evaluate those quantities, only the properties of the particles, here just the effective susceptiblities $\bm{\chi}_i$, and the explicit expression for the vacuum Green's function are required. The vacuum Green's dyadic $\bm{\Gamma}_{0;\zeta}(\mathbf{r},\mathbf{r}')$ is well-known and in this paper they are mainly employed in a $2+1$-dimensional representation, as in Eq.~\eqref{eqRD.p3.1}.

\par Although the theory described here seems quite straightforward, there are still difficulties. Obviously, except for simple cases, practically usable analytic expressions for Casimir free energies, and thus Casimir entropies and forces, are usually out of reach. If taking $k$ nanoparticles into account, the total Casimir free energy $F$, expressed as the sum of $i$-particle interacting contributions, is
\begin{equation}
\label{eqT.8}
F=\sum_{i=1}^kF_i+\sum_{(i_1,i_2)}F_{i_1i_2}+\sum_{(i_1,i_2,i_3)}F_{i_1i_2i_3}
+\cdots+F_{i_1i_2\cdots i_k},
\end{equation}
where $F_{i_1i_2\cdots i_j}$ is the pure $j$-particle interaction free energy of particles ${\rm P}_{i_1},\ {\rm P}_{i_2},\ \cdots,\ {\rm P}_{i_j}$, $(i_1,i_2,\cdots,i_j)$ signifying the labels of these particles. Even though an analytical expression for each term in Eq.~\eqref{eqT.8} can be achieved in an ideal manner, they are typically so complicated that they are not practically and theoretically applicable. In view of this, we resort to the perspective based on the scattering picture, trying to cast some light onto the mechanism,
by which each scattering path contributes to the thermal Casimir interactions among the particles.

\par To simplify the argument and notation, we here state our scheme by considering the zero-temperature case, from which the finite-temperature counterparts can be obtained by
replacing integrals over Euclidean frequencies by sums over Matsubara frequencies. The Casimir free energy at zero temperature, or the Casimir energy, is
\begin{eqnarray}
\label{eqT.9}
U=\frac{1}{2}\int\frac{d\zeta}{2\pi}\Tr\ln(\bm{1}+\bm{\Gamma}_{0;\zeta}
\mathbf{V}_{\zeta}),
\end{eqnarray}
and can be expanded in an explicit form for the spatially non-dispersive materials as
\begin{eqnarray}
\label{eqT.10}
U
&=&
-\frac{1}{2}\sum_{n=1}^{\infty}\frac{(-1)^n}{n}\int\frac{d\zeta}{2\pi}\int d\mathbf{r}_1\cdots d\mathbf{r}_n
\nonumber\\
& &
\times\tr\bigg[
\bm{\Gamma}_{0;\zeta}(\mathbf{r}_n,\mathbf{r}_1)\mathbf{V}_{\zeta}(\mathbf{r}_1)
\bm{\Gamma}_{0;\zeta}(\mathbf{r}_1,\mathbf{r}_2)\mathbf{V}_{\zeta}(\mathbf{r}_2)
\nonumber\\
& &
\cdots\bm{\Gamma}_{0;\zeta}(\mathbf{r}_{n-1},\mathbf{r}_n)\mathbf{V}_{\zeta}(\mathbf{r}_n)
\bigg].
\end{eqnarray}
It is convenient, as in the two- and three-particle cases, to define $\bm{\chi}=\mathbf{V}(\bm{1}+\bm{\Gamma}_0\mathbf{V})^{-1}$, or more explicitly
\begin{eqnarray}
\label{eqT.14}
\bm{\chi}(\zeta;\mathbf{r},\mathbf{r}')
&=&
\mathbf{V}_{\zeta}(\mathbf{r})\delta(\mathbf{r}-\mathbf{r}')
-
\mathbf{V}_{\zeta}(\mathbf{r})\bm{\Gamma}_{0;\zeta}(\mathbf{r},\mathbf{r}')
\mathbf{V}_{\zeta}(\mathbf{r}')
\nonumber\\
& &
-
\mathbf{V}_{\zeta}(\mathbf{r})\sum_{n=3}^{\infty}
\int\prod_{i=1}^{n-2}(-d\mathbf{r}_i)
\bm{\Gamma}_{0;\zeta}(\mathbf{r},\mathbf{r}_1)\mathbf{V}_{\zeta}(\mathbf{r}_1)
\times
\nonumber\\
& &
\cdots\bm{\Gamma}_{0;\zeta}(\mathbf{r}_{n-2},\mathbf{r}')
\mathbf{V}_{\zeta}(\mathbf{r}').
\end{eqnarray}
When there is only one particle in the vacuum, say located at $\bm{0}$, $\bm{\chi}(\zeta;\mathbf{r},\mathbf{r}')$ is
\begin{eqnarray}
\label{eqT.15}
\bm{\chi}(\zeta;\mathbf{r},\mathbf{r}')
=
\bm{\alpha}_{\zeta}\delta(\mathbf{r})\delta(\mathbf{r}-\mathbf{r}')
,
\end{eqnarray}
in which $\bm{\alpha}_{\zeta}$ 
is the polarization; this describes only self-interaction.
For the situation consisting of $N$ particles, $\bm{\chi}(\zeta;\mathbf{r},\mathbf{r}')$ can be expressed in the compact form $\bm{\chi}(\zeta)=\mathbf{V}_{\zeta}(\bm{1}+\bm{\Gamma}_{0;\zeta}\mathbf{V}_{\zeta})^{-1}$ as above, and the bare susceptibility $\mathbf{V}_{\zeta}(\mathbf{r})$ is
\begin{eqnarray}
\label{eqT.16}
\mathbf{V}_{\zeta}(\mathbf{r})
=\sum_{i=1}^{N}\bm{\alpha}^{(0)}_{i}(\zeta)\delta(\mathbf{r}-\mathbf{R}_i)
,
\end{eqnarray}
with $\bm{\alpha}^{(0)}_i$ being the bare susceptibility of the $i$th particle located at $\mathbf{R}_i$. Definitely, the effective susceptibility in Eq.~\eqref{eqT.14} includes the interactions between every part of  system. But we would like to investigate those interactions explicitly, so the renormalized susceptibility of each particle should be utilized. Since the susceptibility of particle in Eq.~\eqref{eqT.14} always exerts its effects in the renormalized form, the effective susceptibility of the $i$th particle, simply denoted as $\bm{\chi}_i=\mathbf{V}_i(\bm{1}+\bm{\Gamma}_{0}\mathbf{V}_i)^{-1}$, is used in the following arguments. (See also Ref.~\cite{guo2021energetics}.) Together with the propagator $-\bm{\Gamma}_0$ connecting two neighboring particles, a scattering path consisting of various particles is specified. Therefore, to label a scattering path, one only needs to know the particles passed through in order on this path. To analyze a scattering path, the details of the path, such as its geometry, should be properly described.

\par To this end, we first bring up a natural frame, based on which we describe the geometry of the path. It is observed that the vacuum Green's dyadic $\bm{\Gamma}_{0;\zeta}(\mathbf{r},\mathbf{r}')$ takes a diagonal form in its principal axis frame, that is,
\begin{equation}
\label{eqRD.p3.1}
\bm{\Gamma}_{0;\zeta}(\mathbf{r},\mathbf{r}')
=
\frac{e^{-d_{\zeta}}}{4\pi d^3}
\bigg[
(\hat{\mathbf{e}}_1\hat{\mathbf{e}}_1+\hat{\mathbf{e}}_2\hat{\mathbf{e}}_2)y_t(d_\zeta) +\hat{\mathbf{e}}_3\hat{\mathbf{e}}_3y_p(d_\zeta)
\bigg]
,
\end{equation}
in which $y_t(x)=x^2+x+1$, $y_p(x)=-2(x+1)$, $d_{\zeta}=d|\zeta|$, $d=|\mathbf{r}-\mathbf{r}'|$, and these three unit vectors satisfy $|\hat{\mathbf{e}}_i|=1$, $\hat{\mathbf{e}}_i\times\hat{\mathbf{e}}_j=\epsilon_{ijk}\hat{\mathbf{e}}_k$, $\hat{\mathbf{e}}_3=(\mathbf{r}-\mathbf{r}')/d$. Note, that we are adopting a $2+1$ spatial decomposition of the Green's dyadic, with the normal direction being given by $\mathbf{\hat{e}}_3$, while the transverse directions are described by $\mathbf{\hat e_{1,2}}$.  This decomposition is natural and convenient, but means that there is a preferred principal axis frame for each propagator, defined by the direction between the two particles the propagator links. In the following arguments, we call the principal axis frame $(\hat{\mathbf{e}}_1,\hat{\mathbf{e}}_2,\hat{\mathbf{e}}_3)$ of the given Green's dyadic $\bm{\Gamma}_0$ as its ``\emph{natural frame}'', and $d$ thus defined as the ``\emph{length}'' of $\bm{\Gamma}_0$. The geometry of a path is uniquely determined by the natural frame and length of each constituent Green's dyadic, or propagator, in the total closed path.

\par To be more specific, consider a closed path passing through $n$ particles (or an \emph{$n$-path}). According to the discussion above, a path $p=[a_1,\cdots,a_n]$ ($a_i$ here labels the particles in the multi-particle system, and $i=1,2,\cdots,n$ signifies the order of particles in this path). The free energy at temperature $T$ is built up of $n$ scattering matrices, as follows
\begin{eqnarray}
\label{eqRD.p3.2}
F_p
&=&
\frac{(-1)^{n+1}m}{n}T\psum_{k=0}^{\infty}
\tr\big[
\bm{\Gamma}_{0;\zeta_k}(\mathbf{r}_{a_n},\mathbf{r}_{a_1})\bm{\alpha}_{a_1}(\zeta_k)
\\
& &
\cdot\bm{\Gamma}_{0;\zeta_k}(\mathbf{r}_{a_1},\mathbf{r}_{a_2})\bm{\alpha}_{a_2}(\zeta_k)\cdots
\bm{\Gamma}_{0;\zeta_k}(\mathbf{r}_{a_{n-1}},\mathbf{r}_{a_n})\bm{\alpha}_{a_n}(\zeta_k)
\big],\nonumber
\end{eqnarray}
with $m$ being the length of circular section of this path\footnote{When a path is given by repeating a section of particles, then we call this section of particles as the circular section of the path. For example, the circular section of the path $[1,2,3,1,2,3]$ is $1,2,3$ and its length is $3$. If the path has no circular section smaller than the total number of particles in the path, then this path itself is its own circular section.}. Note that the particle polarizabilities are not diagonal in any of the natural frames of the Green's dyadics. Denote \emph{the total length of the path} as
\begin{eqnarray}
\label{eqRD.p3.3}
d_{\rm tot}
&=&
d_{n,1}+\sum_{i=2}^{n}d_{i-1,i},\
d_{i,j}=|\mathbf{r}_{a_i}-\mathbf{r}_{a_j}|,
\end{eqnarray}
then the length distribution of this $n$-path is defined as $\bm{\eta}=(\eta_1,\cdots,\eta_n)$, where $\eta_i$ is
\begin{eqnarray}
\label{eqRD.p3.4}
\eta_1=\frac{d_{n,1}}{d_{\rm tot}},\
\eta_i=\frac{d_{i-1,i}}{d_{\rm tot}},\ i=2,\cdots,n;
\end{eqnarray}
$n-1$ independent variables are in $\bm{\eta}$. \emph{The orientation of the path}, represented by the set of natural frames of scattering matrices, are denoted as $\mathcal{O}=(\mathfrak{F}_1,\cdots,\mathfrak{F}_n)$, where $\mathfrak{F}_i$ is the natural principal axis frame of $\bm{\Gamma}_0(\mathbf{r}_{i-1},\mathbf{r}_i)$. To determine the orientation, $2n-6, (n\ge3)$ parameters should be provided, if the rotational symmetry is taken into account, giving a total of $3n-7$ parameters. That is, when $n=2$, apart from the overall scale, no additional parameters are required. For $n=3$ (a triangle), two parameters remain, since two angles
specify the shape.  For $n=4$, the orientation of one more vector is required to specify the tetrahedron in addition, giving a total of 5 parameters.  For each higher polyhedron, one further vector, hence, 3 more parameters, need to be specified. So, apart from the overall scale, $3n-7$ parameters, for $n\ge3$, must be given to describe the configuration.


\par In order to evaluate contributions from a path to the thermal Casimir free energy, it is not enough to only depict its geometry. As noted above, the relative orientation of the frames, which we shall dub channels, 
contained in the path, have to be identified and analyzed as well. With the natural frame utilized, the Green's dyadic $\bm{\Gamma}_{0,\zeta}(\mathbf{r},\mathbf{r}')$ can be organized in a simple matrix format, that is, Eq.~\eqref{eqRD.p3.1} written in terms of ``$(t,p)$'' matrix as (three spatial dimensions are broken into ``2+1'' form, i.e., $t$ signifying two transverse dimensions, $p$ a single longitudinal dimension)
\begin{equation}
\label{eqRD.p3.3a}
\bm{\Gamma}_{0;\zeta}(\mathbf{r},\mathbf{r}')
=
\frac{e^{-d_{\zeta}}}{4\pi d^3}
\left[
  \begin{array}{cc}
    y_t(d_\zeta) & 0 \\
   0  & y_p(d_\zeta) \\
  \end{array}
\right]
,
\end{equation}
in which $y_t$ should be interpreted as the function $y_t$ multiplied by the $2\times2$ identity matrix in this $3\times3$ Green's dyadic. In the same manner, the polarizability of a particle has its $(t,p)$ form as well. It should be noted that, in fact, properties of the polarizability, such as dispersion, anisotropy etc, affect the details of the channels. But we shall not go deeply into the properties of particles for clarity and simplicity in this
paper. For a given particle on the path, the $(t,p)$ form of its polarizability is determined by the natural frames of two Green's dyadics sandwiching it. For example, suppose the natural frames of $\bm{\Gamma}_{0}(\mathbf{r},\mathbf{r}')$ and $\bm{\Gamma}_{0}(\mathbf{r}',\mathbf{r}'')$ are respectively $\hat{\mathbf{e}}_i$ and $\hat{\mathbf{e}}'_i$, then for the expression $\bm{\Gamma}_{0}(\mathbf{r},\mathbf{r}')\bm{\alpha}\bm{\Gamma}_{0}(\mathbf{r}',\mathbf{r}'')$, $\bm{\alpha}$ has the $(t,p)$ form
\begin{equation}
\label{eqRD.p3.4a}
\bm{\alpha}
=
\left[
  \begin{array}{cc}
    \alpha^{tt} & \alpha^{tp} \\
    \alpha^{pt} & \alpha^{pp} \\
  \end{array}
\right]
,
\end{equation}
where $\alpha^{\mu\nu},\ \mu,\nu=t,p$ are blocks of the $3\times3$ polarization matrix, i.e.,
\begin{equation}
\label{eqRD.p3.4b}
\alpha^{tt}
=
\left[
  \begin{array}{cc}
    \alpha_{11} & \alpha_{12} \\
    \alpha_{21} & \alpha_{22} \\
  \end{array}
\right]
,\
\alpha^{tp}
=
\left[
  \begin{array}{c}
    \alpha_{13} \\
    \alpha_{23} \\
  \end{array}
\right]
,\
\alpha^{pt}
=
\left[
  \begin{array}{cc}
    \alpha_{31} & \alpha_{32} \\
  \end{array}
\right]
,\
\end{equation}
and $\alpha^{pp}=\alpha_{33}$, with $\alpha_{ij}=\hat{\mathbf{e}}_i\cdot\bm{\alpha}\cdot\hat{\mathbf{e}}'_j$.
That is, the $\bm{\alpha}$ matrices refer to different frames on the left and right, implying that the polarizability matrix of a particle in this description is not necessarily symmetric. In terms of these $(t,p)$ matrices, $F_p$ is the sum of contributions from $2^n$ channels, and the channel is labeled by the sequence of directions, $t$ and $p$, of out-going propagators. Evidently, channels in a path are largely determined by the polarizabilities of the particles. Influences of polarizabilities are quite complicated, and can be modified manifestly by the orientation of the path.  Unlike in 2-paths, mixing of transverse and longitudinal contributions can be obtained even for an $n>2$-path consisting of only isotropic particles, and one only needs to properly design the geometry of the path, as we see in Sec.~\ref{RD}.

\par By applying this scheme, we explore the thermal aspects of Casimir effects in some relatively simple, yet illustrative, models, attempting to enhance our understanding of the physical origins of negativity, nonmonotonicity and, potentially, divergences of Casimir entropy. The possible nonnegligible impacts from multi-particle interactions, especially the thermal part, in systems with many particles included, are also demonstrated. 

\section{Results and Discussion}
\label{RD}
\par In this section, we evaluate some illustrative configurations and instances, and discuss the thermal corrections to Casimir free energies. Since it is typically quite difficult, or even impossible, to derive  analytical expressions for quantities such as the
Casimir free energy, entropy, force etc, in  multi-particle configurations, we mainly employ
numerical methods to evaluate these quantities. Analytic formulas are given only for simple cases.
\subsection{Two-particle systems}
\label{RD.p2}
\par First, let us investigate the simplest cases, namely systems with only two nanoparticles included. The polarizability of each nanoparticle is assumed to be typically weak, which makes sense considering the small size of the particle~\cite{li2022casimir}; then, the Casimir interaction free energy between two particles, labeled ${\rm P}_1$ and ${\rm P}_2$ as those in Fig.~\ref{figsystem}, can be approximated as
\begin{equation}
\label{eqRD.p2.1}
F_{12}
\approx
-T\psum_{n=0}^{\infty}\Tr
\bm{\Gamma}_{0;\zeta_n}\bm{\alpha}_{1;\zeta_n}
\bm{\Gamma}_{0;\zeta_n}\bm{\alpha}_{2;\zeta_n}
,
\end{equation}
in which only the scattering path passing through the two distinct particles, or 2-path, contributes. In Refs.~[\onlinecite{milton2015negative}] and [\onlinecite{milton2017negative}], the authors studied Casimir-Polder entropies between two particles, but limited to some special cases. Here we generalize this work on two-particle systems. To simplify the evaluation, it is appropriate, without losing any generality, to set the positions of particle $\mathbf{R}_1=\mathbf{0},\ \mathbf{R}_2=Z\hat{\mathbf{z}}$, and the relevant vacuum Green's dyadic satisfies the reciprocity condition $\bm{\Gamma}_{0;\zeta}(\mathbf{R}_1,\mathbf{R}_2)=\bm{\Gamma}_{0;\zeta}(\mathbf{R}_2,\mathbf{R}_1)$ with the following simple matrix form
\begin{eqnarray}
\label{eqRD.p2.2}
\bm{\Gamma}_{0;\zeta}(\mathbf{R}_1,\mathbf{R}_2)
=
\frac{e^{-Z_{\zeta}}}{4\pi|Z|^3}
\begin{small}
\left[
  \begin{array}{ccc}
     y_t(Z_\zeta) & 0 & 0 \\
     0 & y_t(Z_\zeta) & 0 \\
     0 &  0 & y_p(Z_\zeta)  \\
  \end{array}
\right]
\end{small}
,
\end{eqnarray}
in which $Z_\zeta=|Z\zeta|$. Since the $z$-direction of the fixed Cartesian frame is just the principal axis in this case, the evaluation scheme in the $(t,p)$ form follows.

\par Suppose that polarizabilities of these two particles in Eq.~\eqref{eqRD.p2.1} are nondispersive and diagonal. Then the two-particle Casimir entropy, obtained from $F_{12}$, depends on the anisotropy of the particles, that is,
\begin{eqnarray}
\label{eqRD.p2.3}
S_{12}=-\frac{\partial F_{12}}{\partial T}
=
\frac{1}{16\pi^2Z^6}
\bigg[\xi_zs_{z}(Z_T)+\xi_{\perp}s_{\perp}(Z_T)\bigg]
,\quad
\end{eqnarray}
where $Z_T=4\pi T|Z|$,
$\xi_\perp=\xi_x+\xi_y$, $\xi_X=\alpha_{i;X}\alpha_{j;X},\ X=x,y,z$, and the diagonal polarizability of each particle, for instance ${\rm P}_i$, is
\begin{eqnarray}
\label{eqRD.p2.4}
\bm{\alpha}_i=\alpha_{i;x}\mathbf{\hat{x}\hat{x}}+\alpha_{i;y}\mathbf{\hat{y}\hat{y}}
+\alpha_{i;z}\mathbf{\hat{z}\hat{z}},
\end{eqnarray}
in the principal axis frame of the particle, which is the same as the fixed Cartesian frame here. The functions $s_{z}(Z_T)$ and $s_{\perp}(Z_T)$, the explicit definitions of which can be obtained according to Eq.~\eqref{eqRD.p2.1}-\eqref{eqRD.p2.3}, arise from the transverse and longitudinal channels of this 2-path, respectively. Although they have quite complicated analytic forms, their general behaviors are demonstrated schematically in the upper panel of Fig.~\ref{figD01}.
\begin{figure}[h]
  \centering
  \includegraphics[scale=0.36]{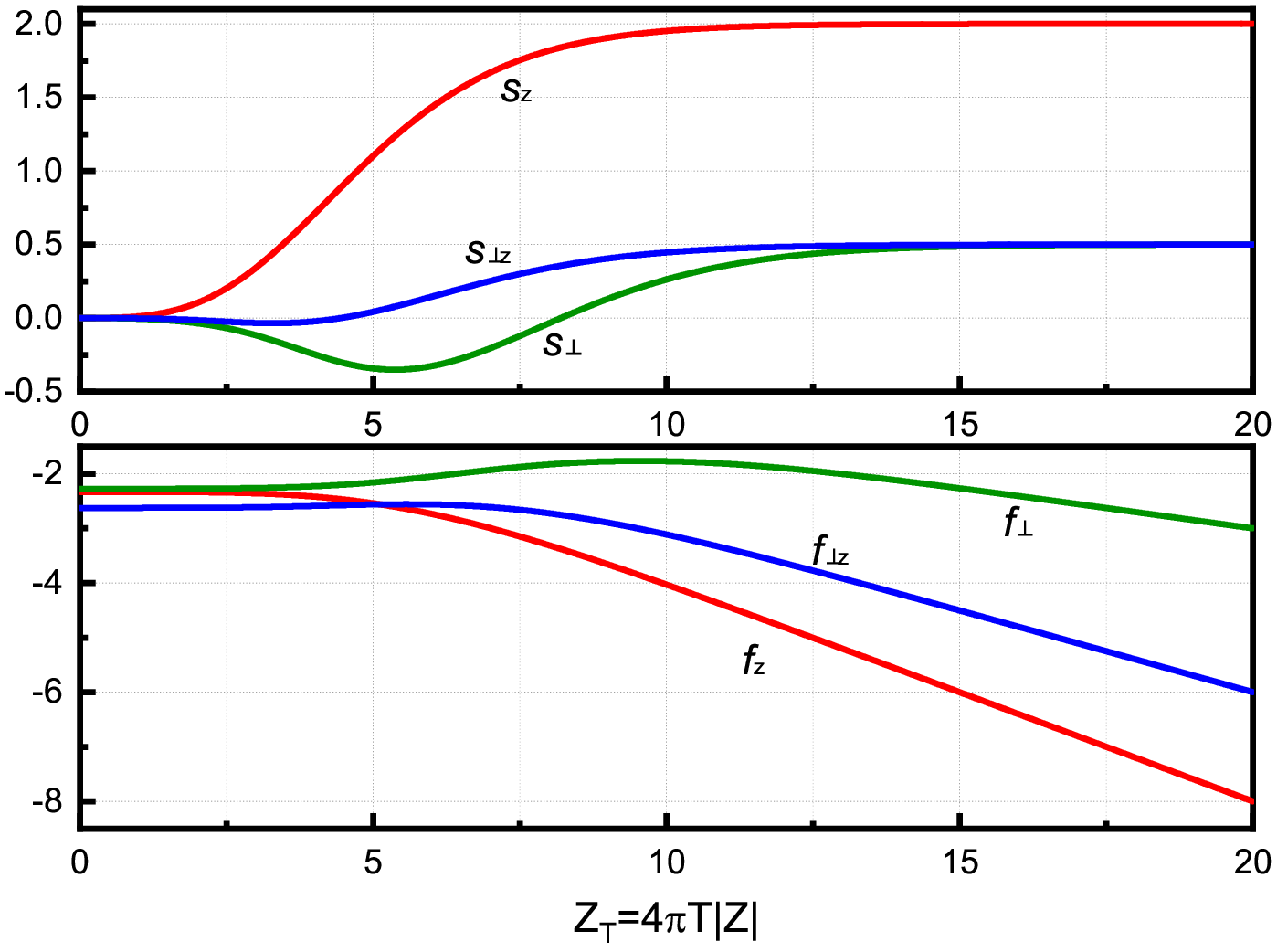}
  \caption{\label{figD01}$s_{z}$ (red), $s_{\perp}$ (green) and $s_{\perp z}$ (blue) (to be defined later) as functions of $Z_T=4\pi T|Z|$ are given in the upper panel, while $f_{z}$ (red), $f_{\perp}$ (green) and $f_{\perp z}$ (blue) as functions of $Z_T=4\pi T|Z|$, divided by $30$, $20$ and $10$ respectively, are given in the lower panel.}
\end{figure}
The longitudinal scattering function $s_z(Z_T)$ is always positive, and approaches a constant monotonically, as the temperature or the separation between the particles increases. In contrast, the transverse scattering function $s_{\perp}(Z_T)$ is non-monotonic, and a negative  region of  $s_{\perp}(Z_T)$ appears, which is the sole origin of the negativity of the Casimir entropy in this system. In the low-temperature limit, $Z_T\ll 1$, we obtain
\begin{subequations}
\label{eqRD.p2.5}
\begin{eqnarray}
\label{eqRD.p2.5a}
s_{z}(Z_T)
\sim
\frac{Z_T^3}{90}+\frac{Z_T^5}{1260}
,\
s_{\perp}(Z_T)
\sim
-\frac{Z_T^3}{180}+\frac{Z_T^5}{1440}
,\quad\quad
\end{eqnarray}
while in the high-temperature or large-distance limit, $Z_T\gg1$, the reduced Casimir entropies settle to the following constants as expected
\begin{eqnarray}
\label{eqRD.p2.5b}
s_{z}(Z_T)
\sim
2
,\
s_{\perp}(Z_T)
\sim
\frac{1}{2}
.
\end{eqnarray}
\end{subequations}

\par In  previous work~\cite{milton2015negative,milton2017negative}, we have studied the Casimir-Polder interaction between two particles with diagonal nondispersive electric polarizabilities. To make comparison with those results, we evaluate the case in which the polarizabilities of ${\rm P}_1$ and ${\rm P}_2$, $\bm{\alpha}_i,\ i=1,2$, satisfy $\alpha_{1;z}\alpha_{2;z}\neq0$ and $\bm{\alpha}_i=\diag(\alpha_{i;\perp},\alpha_{i;\perp},\alpha_{i;z})$. Then the Casimir entropy is
\begin{eqnarray}
\label{eqRD.p2.6}
S_{12}
=
\frac{\alpha_{1;z}\alpha_{2;z}}{16\pi^2Z^6}
\bigg[s_{z}(Z_T)+2\gamma s_{\perp}(Z_T)\bigg]
,
\end{eqnarray}
where the anisotropic index is defined as $\gamma=\gamma_1\gamma_2,\ \gamma_i=\alpha_{i;\perp}/\alpha_{i;z}$. Limiting behaviors are obtained directly from Eq.~\eqref{eqRD.p2.5}, as
\begin{eqnarray}
\label{eqRD.p2.7}
&&
Z_T\rightarrow0:\quad
S_{12}
\sim
\frac{\alpha_{1;z}\alpha_{2;z}}{16\pi^2Z^6}
\bigg(\frac{1-\gamma}{90}Z_T^3+\frac{4+7\gamma}{5040}Z_T^5\bigg)
,
\nonumber\\
&&
Z_T\rightarrow\infty:\quad
S_{12}
\sim
\frac{\alpha_{1;z}\alpha_{2;z}}{16\pi^2Z^6}
(2+\gamma),
\end{eqnarray}
which are consistent with Eq.~(3.6) of Ref.~[\onlinecite{milton2015negative}] or
Eq.~(17) in Ref.~[\onlinecite{milton2017negative}]. If $0<\gamma<1$, $S_{12}$ can never be negative, since the positive longitudinal contribution always overwhelms the negative transverse one. If $\gamma>1$, the larger $\gamma$ becomes (implying that these two particles are more anisotropic), the more negative this entropy can be. But when the interparticle distance or the temperature is sufficiently large, the contributions from both types of scattering are positive, no matter how anisotropic the particles are.
\begin{figure*}
  \centering
  \includegraphics[scale=1.045]{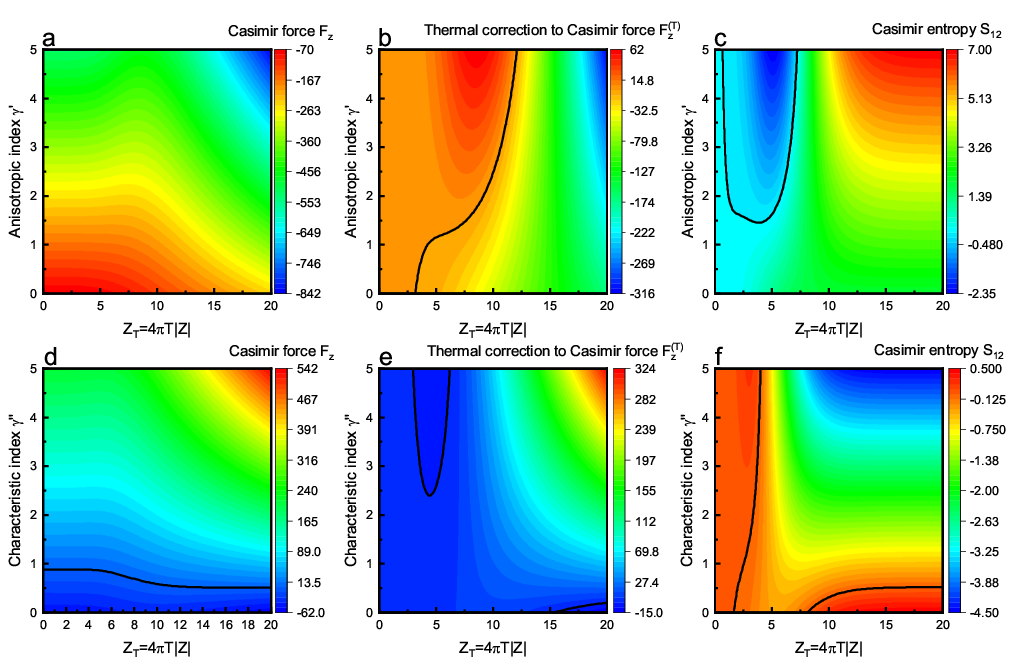}
  \caption{\label{figA22}\textbf{(a)} and \textbf{(b)} demonstrate the Casimir force $F_z$ in Eq.~\eqref{eqRD.p2.8} and the thermal correction to the Casimir force $F_z^{(T)}$ in Eq.~\eqref{eqRD.p2.12} (in units of $\alpha_{1;z}\alpha_{1;z}/64\pi^3Z^8$) as functions of $Z_T=4\pi T|Z|$ and anisotropic index $\gamma'$. The Casimir entropy $S_{12}$ in Eq.~\eqref{eqRD.p2.3} (in units of $\alpha_{1;z}\alpha_{1;z}/16\pi^2Z^6$) as a function of $Z_T=4\pi T|Z|$ and anisotropic index $\gamma'$, is shown in \textbf{(c)}. \textbf{(d)} and \textbf{(e)} show the Casimir force $F_z$ in Eq.~\eqref{eqRD.p2.19} and its thermal correction $F_z^{(T)}$ (in units of $\chi_{xy}/64\pi^3Z^8$) as functions of $Z_T=4\pi T|Z|$ and characteristic index $\gamma''$, while \textbf{(f)} shows the Casimir entropy $S_{12}$ in Eq.~\eqref{eqRD.p2.17} (in units of $\chi_{xy}/16\pi^2Z^6$) as a function of $Z_T=4\pi T|Z|$ and characteristic index $\gamma''$. The zero lines are given to guide
  the eye.  
  }
\end{figure*}

\par It should be mentioned that the appearance of a negative entropy means physically we are dealing with a part of the whole system. For the whole system, when all contributions, including those from the various channels, are taken into account, the total entropy has to be positive. The entropy of the electromagnetic vacuum will overwhelm the negative Casimir entropy, rendering the total entropy positive. Although no thermodynamic law is violated, the negativity and nonmonotonicity of Casimir entropy imply nontrivial properties.

\par It is helpful to understand the relatively abstract Casimir entropy, which embodies the properties of thermal corrections to the zero-temperature Casimir free energy, by connecting it to some more experimentally accessible quantities, such as the Casimir force. In the two-particle system above, the Casimir-Polder force acting on the particle is in the $z$-direction and can be expressed, in terms of the free energy, as
\begin{eqnarray}
\label{eqRD.p2.8}
\mathcal{F}_z
=
-\frac{\partial F_{12}
}{
\partial|Z|}
=
\frac{1}{64\pi^3Z^8}
\bigg[\xi_zf_{z}(Z_T)+\xi_{\perp}f_{\perp}(Z_T)\bigg]
.\quad
\end{eqnarray}
Similarly to the entropy, the analytic expressions for $f_z$ and $f_{\perp}$, due to the longitudinal and transverse scattering channels respectively, are complicated, while their general behaviors are demonstrated schematically in the lower panel of Fig.~\ref{figD01}. In the low-temperature limit $Z_T\ll1$, $f_{z}(Z_T)$ and $f_{\perp}(Z_T)$ are approximated by
\begin{eqnarray}
\label{eqRD.p2.9}
f_{z}(Z_T)\sim
-70-\frac{Z_T^4}{120}
,\
f_{\perp}(Z_T)\sim
-\frac{91}{2}+\frac{Z_T^4}{240}
,
\end{eqnarray}
which means for isotropic particles ($2\chi_z=\chi_\perp$), the famous retarded van der Waals (Casimir-Polder) force at zero temperature is recovered. Although in the zero-temperature case, the force from each channel is always attractive, channels do not contribute to the force equally. Suppose $\alpha_{1;z}\alpha_{2;z}\neq0$, then for small $|Z|T$, $\mathcal{F}_z$ can be written as
\begin{eqnarray}
\label{eqRD.p2.10}
\mathcal{F}_z
&\approx&
\frac{\alpha_{1;z}\alpha_{2;z}}{64\pi^3Z^8}
\bigg(
-70-91\gamma'+\frac{\gamma'-1}{120}Z_T^4
\bigg)
,
\end{eqnarray}
where the general anisotropic index is defined as $\gamma'=(\alpha_{1;x}\alpha_{2;x}+\alpha_{1;y}\alpha_{2;y})/2\alpha_{1;z}\alpha_{2;z}$. When $\gamma'\neq1$ holds, then the thermal correction to the force may be either attractive or repulsive, depending on details of the anisotropy.
In the high-temperature or large-distance limit $Z_T\gg1$, the reduced Casimir-Polder forces $f_{z}(Z_T)$ and $f_{\perp}(Z_T)$ behave asymptotically as
\begin{eqnarray}
\label{eqRD.p2.11}
f_{z}(Z_T)\sim
-12Z_T
,\
f_{\perp}(Z_T)\sim
-3Z_T
,
\end{eqnarray}
which are consistent with the isotropic results given in Eq.~(A1) of Ref.~\cite{li2022casimir} and further consistent with the
entropy limit found in Eq.~(\ref{eqRD.p2.7})--see Eq.~(\ref{eqRD.p2.13}) below. According to the lower panel of Fig.~\ref{figD01}, the non-monotonicity of Casimir force exhibits correspondences with that of Casimir entropy, even with the zero-temperature contributions still included. In light of the view that the Casimir entropy is the quantity best illustrating the thermal corrections, we can explicitly extract the thermal part of the Casimir force as follows:
\begin{equation}
\label{eqRD.p2.12}
\mathcal{F}_z^{(T)}
=
\frac{\alpha_{1;z}\alpha_{2;z}}{64\pi^3Z^8}f_z^{(T)}(\gamma',Z_T)
,
\end{equation}
where the reduced thermal correction $f_z^{(T)}(\gamma',Z_T)$ is
\begin{equation}
\label{eqRD.p2.13}
f_z^{(T)}(\gamma',Z_T)=
Z_Ts_{12}(\gamma',Z_T)
-
7\int^{Z_T}_0dx
s_{12}(\gamma',x)
,
\end{equation}
and $s_{12}(\gamma',x)=s_{z}(x)+2\gamma's_{\perp}(x)$ is the reduced Casimir entropy of these two particles, the general behaviors of which are shown by Fig.~\ref{figA22}c. According to the general behaviors of $\mathcal{F}_z$ in Eq.~\eqref{eqRD.p2.8} described by Fig.~\ref{figA22}a, the force between these two particles is always attractive, but its dependence on the temperature is not monotonic. With a larger anisotropic index, the the magnitude of $\mathcal{F}_{z}$ typically increases. If $\gamma'$ is large enough,
$\mathcal{F}_{z}$ exhibits
nonmonotonic dependence on the temperature. This nonmonotonicity in both Casimir entropy and force is strengthened by the anisotropy. Furthermore, the thermal correction to the Casimir force $\mathcal{F}_z^{(T)}$ in Eq.~\eqref{eqRD.p2.12}, behaves nonmonotonically, exhibiting a clear similarity with that of the Casimir entropy $S_{12}$ shown in Fig.~\ref{figA22}c. The influences on the thermal correction are evident, and  anisotropy can lead to the negativity of the Casimir entropy and the nonmonotonicity of the Casimir force, by amplifying or reducing the impacts of transverse channels. Obviously, $f^{(T)}_z$ and thus $\mathcal{F}^{(T)}_z$ vanish with $T\rightarrow0$, and according to Eq.~(\ref{eqRD.p2.10}), $f^{(T)}_z$ near $Z_T=0$ satisfies
\begin{equation}
\label{eqRD.p2.14}
\frac{\partial f_z^{(T)}(\gamma',Z_T)}{\partial Z_T}\approx
\frac{\gamma'-1}{30}Z_T^3
,
\end{equation}
which means for $\gamma'>1$, the thermal correction will be positive for the relatively low temperature. As the temperature rises, $\mathcal{F}^{(T)}_z$ will finally turn negative, if $\gamma'>1$ according to Eq.~\eqref{eqRD.p2.11}. Therefore, the dependence of the thermal correction on the temperature will be nonmonotonic at least for $\gamma'>1$.

\par In the model above, transverse and longitudinal scatterings channels do not mix. For particles with off-diagonal polarizabilities (to emphasize the nonmonotonicity and negativity,
here let us ignore the diagonal terms, since the diagonal terms and off-diagonal terms do not interfere within the $2$-path written in Eq.~\eqref{eqRD.p2.1}), there are more subtleties. The Casimir entropy is now written as
\begin{eqnarray}
\label{eqRD.p2.15}
S_{12}
=
\frac{1}{16\pi^2Z^6}
\bigg[
\xi_{xy}s_{\perp}(Z_T)
-
2\xi_{\perp z}s_{\perp z}(Z_T)
\bigg]
,
\end{eqnarray}
where $\xi_{xy}=\alpha_{1;xy}\alpha_{2;yx}+\alpha_{1;yx}\alpha_{2;xy},\ \xi_{\perp z}=\alpha_{1;xz}\alpha_{2;zx}+\alpha_{1;yz}\alpha_{2;zy}+\alpha_{1;zx}\alpha_{2;xz}
+\alpha_{1;zy}\alpha_{2;yz}$.
In Eq.~\eqref{eqRD.p2.15}, the transverse channel remains, but the mixing channel, instead of the longitudinal one, occurs. The mixed reduced entropy, $s_{\perp z}$, corresponding to the mixing channel, is depicted in Fig.~\ref{figD01}, and its limiting behaviors are
\begin{eqnarray}
\label{eqRD.p2.16}
&&
Z_T\rightarrow0:\quad
s_{\perp z}(Z_T)\sim-\frac{Z_T^3}{720}-\frac{Z_T^5}{3360}
,\nonumber\\
&&
Z_T\rightarrow\infty:\quad
s_{\perp z}(Z_T)\sim\frac{1}{2}.
\end{eqnarray}
Suppose $\alpha_{i;xy}\alpha_{j;yx}\ne0$, thus $\xi_{xy}$, is nonzero, then Eq.~\eqref{eqRD.p2.15} is
\begin{eqnarray}
\label{eqRD.p2.17}
S_{12}
=
\frac{\xi_{xy}}{16\pi^2Z^6}
\bigg[
s_{\perp}(Z_T)
-
2\gamma''s_{\perp z}(Z_T)
\bigg]
,
\end{eqnarray}
where the characteristic index is $\gamma''=\xi_{\perp z}/\xi_{xy}$, indicating the relative strength between transverse and mixing channels. The low-temperature limiting behaviors of $S_{12}$ are shown as
\begin{subequations}
\label{eqRD.p2.18}
\begin{eqnarray}
\label{eqRD.p2.18a}
S_{12}
\sim
\frac{\xi_{xy}}{16\pi^2Z^6}
\bigg[
-\frac{2-\gamma''}{360}Z_T^3+\frac{7+6\gamma''}{10080}Z_T^5
\bigg]
,\quad
\end{eqnarray}
while in the large-distance or high-temperature limit, $S_{12}$ behaves as
\begin{eqnarray}
\label{eqRD.p2.18b}
S_{12}
\sim
\frac{\xi_{xy}}{16\pi^2Z^6}
\frac{1-2\gamma''}{2}
.
\end{eqnarray}
\end{subequations}
Similarly, the Casimir force acting on the particle is in the $z$-direction and expressed as
\begin{eqnarray}
\label{eqRD.p2.19}
\mathcal{F}_z
&=&
\frac{1}{64\pi^3Z^8}
\bigg[\xi_{xy}f_{\perp}(Z_T)
-
2\xi_{\perp z}f_{\perp z}(Z_T)\bigg]
\nonumber\\
&=&
\frac{\xi_{xy}}{64\pi^3Z^8}
\bigg[f_{\perp}(Z_T)
-
2\gamma''f_{\perp z}(Z_T)\bigg]
,
\end{eqnarray}
in the second step of which $\xi_{xy}$ is assumed nonzero as in Eq.~\eqref{eqRD.p2.17}. The limiting behaviors of $\mathcal{F}_z$ are simply
\begin{subequations}
\label{eqRD.p2.20}
\begin{equation}
\label{eqRD.p2.20a}
\mathcal{F}_z
\sim
\frac{1}{64\pi^3Z^8}
\bigg(
\frac{105\xi_{\perp z}-91\xi_{xy}}{2}+\frac{2\xi_{xy}-\xi_{\perp z}}{480}Z_T^4
\bigg)
,\quad\quad
\end{equation}
for $Z_T\ll1$, and for $Z_T\gg1$
\begin{eqnarray}
\label{eqRD.p2.20b}
\mathcal{F}_z
=
\frac{3(2\xi_{\perp z}-\xi_{xy})}{64\pi^3Z^8}Z_T
.
\end{eqnarray}
\end{subequations}
The thermal part of the Casimir force $\mathcal{F}_z$ in Eq.~\eqref{eqRD.p2.19} can be obtained via the corresponding Casimir entropy analogous to Eq.~\eqref{eqRD.p2.12} and Eq.~\eqref{eqRD.p2.13}. The general behaviors of $S_{12}$ in Eq.~\eqref{eqRD.p2.17} are shown by Fig.~\ref{figA22}f. The sign of off-diagonal Casimir entropy is highly dependent on the parameter $\gamma''$, and according to Eq.~\eqref{eqRD.p2.18}, $S_{12}$, induced by off-diagonal contributions, for fixed $\gamma''$, always has a region of $T$ for which it is negative.
Fig.~\ref{figA22}f also demonstrates the nonmonotonicity of this $S_{12}$, especially for small $\gamma''$, which is consistent with the results in Fig.~\ref{figD01}. The Casimir force due to the off-diagonal polarizations can turn repulsive for most temperatures $T$ and characteristic indices $\gamma''$, in contrast to the diagonal Casimir force described by Eq.~\eqref{eqRD.p2.8}. Thermal corrections generally contribute significantly to the Casimir forces when the temperature is high or the separation is large, in both the off-diagonal cases and diagonal cases. The correspondences between the Casimir entropies and thermal corrections to the force are evident. The transverse and mixing channels, especially the mixing one, can serve as the source of nonmonotonicity and negativity of Casimir entropy. But the vital impacts due to the geometry of the path is not applicable in two-particle cases.

\subsection{Three-particle systems\footnote{The earliest work on three-body van der Waals interations, which are nonretarded Casimir-Polder interactions, was by Axilrod and Teller
\cite{Axilrod}. They only considered zeo-temperature configurations, so the connection with our work is rather remote.}}
\label{RD.p3}

\par As revealed by arguments on two-particle configurations, one of the main effects of polarizabilities is to select the relevant scattering channels between particles. In other words, polarizabilities of particles determine the channels included in a scattering path, as shown explicitly in Sec.~\ref{RD.p2}. Nevertheless, within multi-particle systems, it is not straightforward to specify the contributions from each channel as in the two-particle cases. First, the geometry of the path, by itself, can induce the mixing of transverse and longitudinal scatterings. Second, extra complexities are largely introduced by the diversity of paths in multi-particle systems.
\par As an illuminating example, let us study  3-path configurations, where three different particles are involved, as depicted schematically by Fig.~\ref{figsystem}. Without losing any generality, we focus on the path $\mathfrak{P}=[1,2,3]$. According to Eq.~\eqref{eqT.4}, as well as the picture of the fluctuating field, the Casimir free energy, owing to the three-particle interaction, is
\begin{eqnarray}
\label{eqRD.p3.5}
F_{\mathfrak{P}}
&=&
T\psum_{n=0}^{\infty}\Tr
\bigg[
\bm{\Gamma}_{0;\zeta_n}(\mathbf{R}_3,\mathbf{R}_1)\bm{\alpha}_{1}(\zeta_n)
\bm{\Gamma}_{0;\zeta_n}(\mathbf{R}_1,\mathbf{R}_2)
\nonumber\\
& &
\times\bm{\alpha}_{2}(\zeta_n)
\bm{\Gamma}_{0;\zeta_n}(\mathbf{R}_2,\mathbf{R}_3)\bm{\alpha}_{3}(\zeta_n)
\bigg]
,
\end{eqnarray}
with $\mathbf{R}_i$ and $\bm{\alpha}_{i}$ being the location and the  polarizability of ${\rm P}_i$. As in Fig.~\ref{figsystem}, we set $\mathbf{R}_1=a\hat{\mathbf{z}}=-\mathbf{R}_2$. For a three-particle system, particles are always located in the same plane, and given the symmetry, it is enough to assume ${\rm P}_3$ is located in the half-infinite plane, which contains the nonnegative $y$-axis and perpendicular to $x$-axis, that is, we choose $\phi=\pi/2$. The geometric configuration of this system can be determined by two parameters as predicted in Sec.~\ref{T}, i.e. the polar angle $\theta$ and ratio of distances $r=|\mathbf{R}_3|/a$ of ${\rm P}_3$ in Fig.~\ref{figsystem}. The explicit formula of $F_{\mathfrak{P}}$ in Eq.~\eqref{eqRD.p3.5} is typically complex. To demonstrate the influences of geometry clearly, however, we here only investigate the nondispersive isotropic particles and consider four representative channels, namely $ttt,\ tpt,\ tpp$ and $ppp$, in this path. Suppose each particle has the polarizability of the form $\bm{\alpha}_i=\alpha_i\bm{1}$, then the free energy from the channel $\mu\nu\lambda,\ \mu,\nu,\lambda=t,p$ is simply
\begin{eqnarray}
\label{eqRD.p3.6}
F_{\mathfrak{P}}^{\mu\nu\lambda}
&=&
\frac{\alpha_1\alpha_2\alpha_3}{128\pi^4d^{10}}\frac{\gamma_{\mu\nu\lambda}
}{\eta_1^3\eta_2^3\eta_3^3}
f_{\mu\nu\lambda}(\bm{\eta};d_T)
,
\end{eqnarray}
where $d$ is the length of the path, $d_T=2\pi Td$. Here, the direction cosines between the frames are simply given in terms of a trace over projection operators:
\begin{equation}
\gamma_{\mu\nu\lambda}=\tr\pi_1^\mu\pi_2^\nu\pi_3^\lambda,\quad \mu,\nu,\lambda\in(t,p),
\end{equation} where the $p$ and $t$ projection operators for each frame are
\begin{equation}
\pi_i^p=\hat{\mathbf{e}}_{3}^i\hat{\mathbf{e}}_3^i,\quad
\pi_i^t=\hat{\mathbf{e}}_1^i\hat{\mathbf{e}}_1^i
+\hat{\mathbf{e}}_2^i\hat{\mathbf{e}}_2^i.
\end{equation}
The function $f_{\mu\nu\lambda}(\bm{\eta};d_T)$ detailed below is independent of the electric response of each particle because the latter are assumed nondispersive:
\begin{eqnarray}
\label{eqRD.p3.6b}
f_{\mu\nu\lambda}(\bm{\eta};d_T)
&=&
d_{T}
\psum_{n=0}^{\infty}y_{\mu}(\eta_1d_{T}n)
y_{\nu}(\eta_2d_{T}n)
\nonumber\\
& &
\times y_{\lambda}(\eta_3d_{T}n)e^{-d_{T}n}
.
\end{eqnarray}
Evidently, when the geometry of the path and polarizabilities of particles are fixed, the Casimir energies, obtained at zero temperature, induced by those four channels  all scale in $d$ as $O(d^{-10})$. Since another scale $1/2\pi T$ is present if the temperature is finite, the scaling law significantly deviates from its zero-temperature counterpart. The low- and high-temperature behaviors of the Casimir entropy, stemming from those four channels, are listed in Table.~\ref{tb1}. The low-temperature scaling law in $d$ of the entropy is $O(d^{-6})$, while the high-temperature one is $O(d^{-9})$.
\begin{table}
\centering
\begin{tabular}{c|cc}
  \hline
   Channels & Low-$T$ & High-$T$ \\
  \hline
   $ttt$ & $\frac{1-3(1-\eta_1)(1-\eta_2)(1-\eta_3)}{45}d_T^3$ & $-\frac{1}{2}$ \\
   $tpt$ & $\frac{1-3(\eta_1^2+\eta_2+\eta_3^2)(1-\eta_2)}{45}d_T^3$ & $1$ \\
   $tpp$ & $\frac{6\eta_1^3+6(\eta_1\eta_2+\eta_1\eta_3+\eta_2\eta_3-\eta_1\eta_2\eta_3)-2}{45}d_T^3$ & $-2$ \\
   $ppp$ & $\frac{4-12(\eta_1\eta_2+\eta_1\eta_3+\eta_2\eta_3-\eta_1\eta_2\eta_3)}{45}d_T^3$ & $4$ \\
  \hline
\end{tabular}
\caption{\label{tb1}The low- and high-temperature limits of the Casimir entropies for given channels $-\partial F^{\mu\nu\lambda}_p/\partial T$ according to Eq.~\eqref{eqRD.p3.6}, in units of $\alpha_1\alpha_2\alpha_3\gamma_{\mu\nu\lambda}/64\pi^3d^{9}\eta_1^3\eta_2^3\eta_3^3$. The low- and high-temperature limits are readily obtained directly from the sum in Eq.~\eqref{eqRD.p3.6b}; the former can also be verified through the Euler-Maclaurin formula.}
\end{table}
Besides the dependence on the length of the path, the temperature-dependent part is also affected by the length distribution $\bm{\eta}$. The low-temperature limiting results from channels $ttt$ and $ppp$, in Table.~\ref{tb1}, are always positive, while the high-temperature limits are, respectively, negative and positive. For $ttt$, a sign change occurs in the Casimir entropy with increasing temperature, while the $ppp$ contribution apparently does not change sign. As for mixing channels $tpt$ and $tpp$, their low-temperature limits can be either positive or negative, depending on the geometry of the path. Negative entropy contributions in paths including more than two scattering matrices are hence commonly seen. To give a first indication of the dependence on temperature of these four channels, we evaluate the configuration with the ratio of distances of ${\rm P}_3$ $r=1$ and the polar angle $\theta=\pi/2$, shown in Fig.~\ref{fig3PNondisp_temperature}. The nonmonotonicity is clearly seen in the behaviors of $ttt$ and $tpt$ channels, which are qualitatively different from the $tpp$ and $ppp$ channels. Actually the sign flipping of Casimir entropy is closely related to the nonmonotonicity of the free energy, which is also strongly affected by the geometry of the path.
\begin{figure}
  \centering
  \includegraphics[scale=0.40]{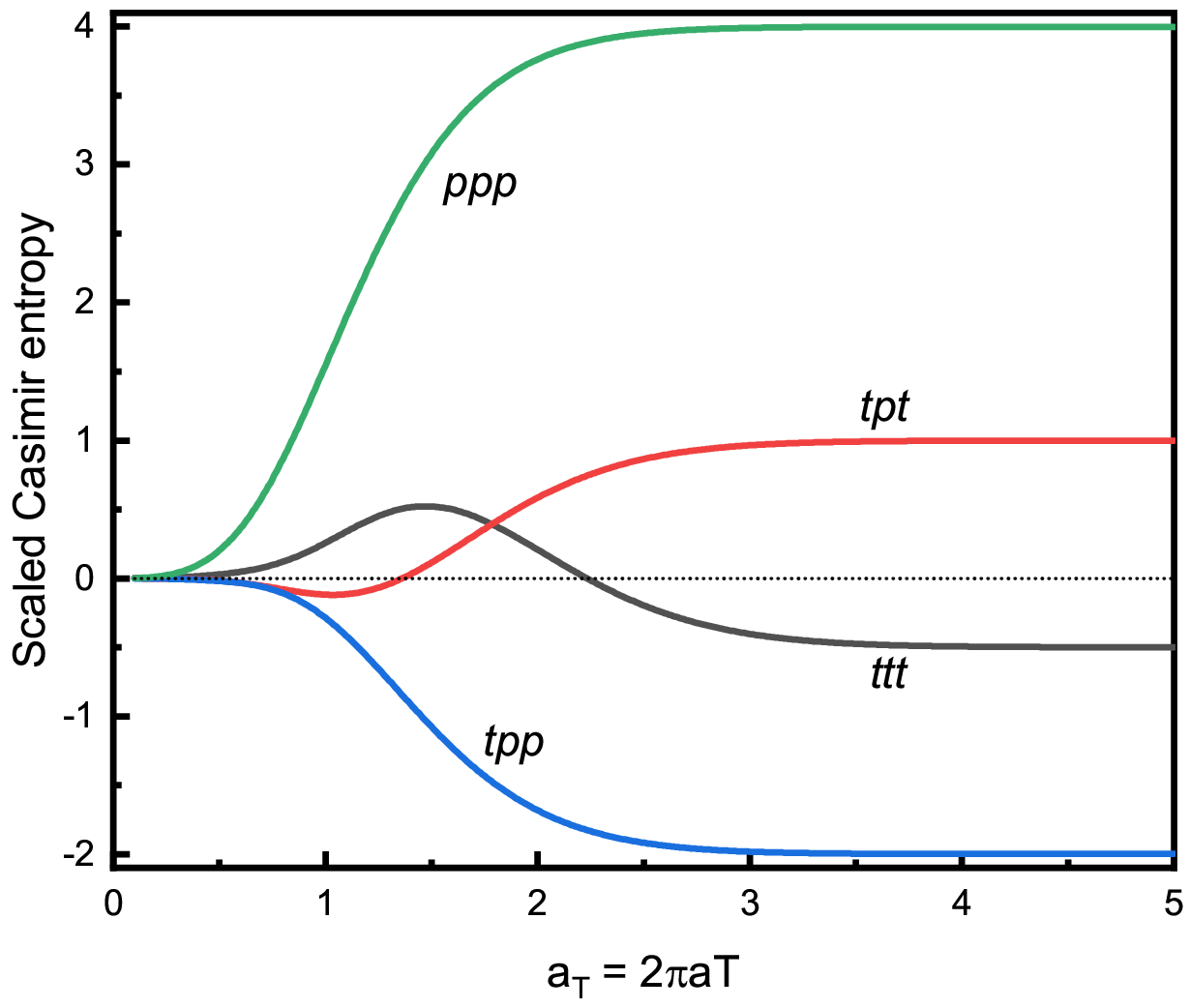}
  \caption{\label{fig3PNondisp_temperature}The Casimir entropies induced by channels $ttt$ (black), $tpt$ (red), $tpp$ (blue) and $ppp$ (green), in units of $\alpha_1\alpha_2\alpha_3\gamma_{\mu\nu\lambda}/64\pi^3d^{9}\eta_1^3\eta_2^3\eta_3^3$, as functions of $a_T=2\pi aT$ with $a$ fixed. The geometry of the path is characterized by the ratio of distances of ${\rm P}_3$ being $r=1$ and the polar angle being $\theta=\pi/2$.}
\end{figure}

\begin{figure*}
  \centering
  \includegraphics[scale=0.9]{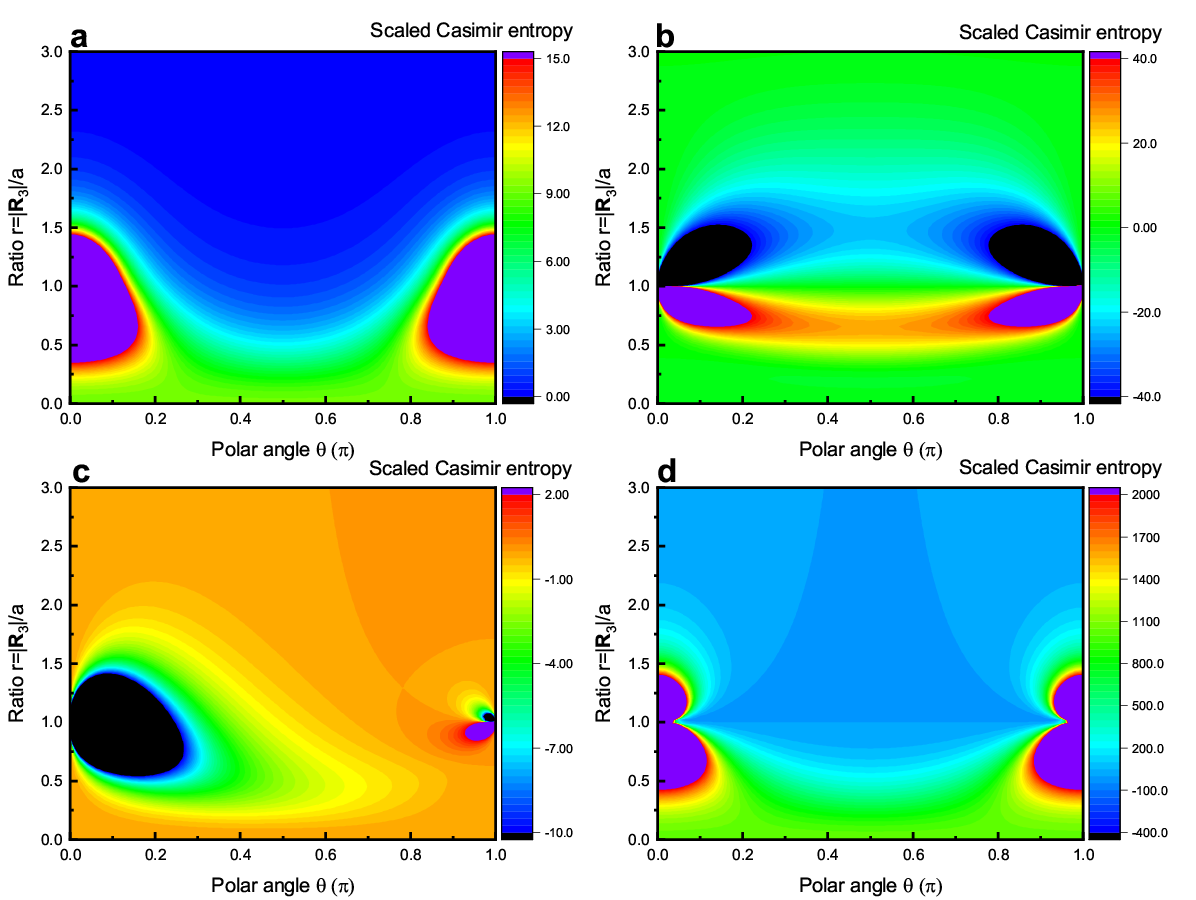}
  \caption{\label{fig3PNondisp_geometry}The Casimir entropies induced by channels $ttt$ \textbf{(a)}, $tpt$ \textbf{(b)}, $tpp$ \textbf{(c)} and $ppp$ \textbf{(d)}, as functions of the ratio of distance $r=|\mathbf{R}_3|/a$ and the polar angle $\theta$. The results for each channel is scaled by the absolute value of its Casimir entropy with the parameters $r=1,\ \theta=\pi/2$, and $2\pi a T=1$.}
\end{figure*}
\par To address the effects of the geometry specifically, especially the orientation of the path, we numerically evaluate the Casimir entropies from those four channels above, with fixed $a$ and $T$ satisfying $a_T=2\pi aT=1$, depicted in Fig.~\ref{fig3PNondisp_geometry}, where the nonmonotonicity and negativity of Casimir entropy are displayed conspicuously. In Fig.~\ref{fig3PNondisp_geometry}a, b and d, we observe symmetry with respect to the polar angle $\theta$. Without the longitudinal scattering involved in this channel, the Casimir entropy is always positive in the parameter region, as shown in Fig.~\ref{fig3PNondisp_geometry}a. Otherwise,  negative Casimir entropy is allowed in a relatively large parameter region, indicating that the negativity of Casimir entropy is almost inevitable in the contributions from many-body channels. The channel $tpp$, in Fig.~\ref{fig3PNondisp_geometry}c, exhibits a strong asymmetry as anticipated. Its Casimir entropy is typically negative, and a deeply negative region is found close to $r=1,\ \theta=0$, that is, the location of ${\rm P}_1$, while around ${\rm P}_2$ there exist tiny positive and negative regions. It is noteworthy that the orientation of the path, reflected in the direction cosine
factor $\gamma_{\mu\nu\lambda}$, illustrates its impacts in the peculiar behaviors of Casimir entropy near vicinities of ${\rm P}_1$ and ${\rm P}_2$. For $ttt,\ tpt$ and $ppp$, it is sufficient to just evaluate the case where ${\rm P}_3$ is near ${\rm P}_1$. Algebraically, the geometric factors $\gamma_{\mu\nu\lambda}$ for each channel are given by the following, where we also show the behavior as
 ${\rm P}_3$ approaches ${\rm P}_1$,
\begin{subequations}
\label{eqRD.p3.7}
\begin{eqnarray}
\label{eqRD.p3.7a}
\gamma_{ttt}=1-\cos\theta_{23}\cos(\theta_{23}+\theta_{31})\cos\theta_{31}\sim O(\eta_1^0),\quad
\end{eqnarray}
\begin{eqnarray}
\label{eqRD.p3.7b}
\gamma_{tpt}=-\sin\theta_{23}\cos(\theta_{23}+\theta_{31})\sin\theta_{31}\sim O(\eta_1),\quad
\end{eqnarray}
\begin{eqnarray}
\label{eqRD.p3.7c}
\gamma_{ppp}=\cos\theta_{23}\cos(\theta_{23}+\theta_{31})\cos\theta_{31}\sim O(\eta_1^0).\quad
\end{eqnarray}
Here $\theta_{ij}$ is the angle between the vector $\mathbf{R}_i-\mathbf{R}_j$ of $\bm{\Gamma}_{0;\zeta_n}(\mathbf{R}_i,\mathbf{R}_j)$ and $z$-axis, and the order of the dependence on the relative distance between ${\rm P}_3$ and ${\rm P}_1$, $\eta_1$, is provided. These formulas qualitatively clarify the results in Fig.~\ref{fig3PNondisp_geometry}a, b and d. For the $tpp$ channel, $\gamma_{tpp}$ 
behaves differently near ${\rm P}_1$ and ${\rm P}_2$ as we see from
\begin{eqnarray}
\label{eqRD.p3.7d}
 \gamma_{tpp}&=&\cos\theta_{23}\sin(\theta_{23}+\theta_{31})\sin\theta_{31}\nonumber\\
 &\sim& O(\eta_1^0)\ \mbox{near ${\rm P}_1$},\quad \sim O(\eta_3) \ \mbox{near ${\rm P}_2$},
\end{eqnarray}
\end{subequations}
resulting in the asymmetry seen in Fig.~\ref{fig3PNondisp_geometry}c. Together with the length distribution $\bm{\eta}$, giving the divergences $\eta_1^{-3}$ near $\rm P_1$ and $\eta_3^{-3}$ near $\rm P_2$, the results in Fig.~\ref{fig3PNondisp_geometry} are justified, and the significant influences due to the geometry have been explicitly manifested.

\par In view of the results and arguments about the simple model above, it can be claimed that the geometry of the path, as well as the properties of polarizabilities, would introduce a huge number of degrees of freedom, if one attempts to describe the Casimir interactions in a system consisting of various particles with nondiagonal polarizabilities, and the thermal corrections to these interactions would be diverse and subtle. In fact, researchers have already tried to unveil the properties of multi-body systems, and it is helpful to review the thermal Casimir interactions in those systems briefly here. As an illustration, consider the simplest three-body interacting system in Ref.~[\onlinecite{milton2013three}], that is, two isotropically polarizable atoms, equidistant from a perfectly conducting plate. By introducing images of these two particles as shown in the inset of Fig.~\ref{figsystem} based on the boundary condition on a perfectly conducting mirror, we
see that contributions to the Casimir free energy from 2-paths include not only that due to the direct interaction between the two particles $F_{12}$, but also paths involving one of the images, $F_{1\bar{2}},\ F_{\bar{1}2}$, or both $F_{\bar{1}\bar{2}}$. The effects of the perfectly conducting mirror are caused by the extra paths with various geometries. At zero temperature, $F_{1\bar{2}},\ F_{\bar{1}2}$ and $F_{\bar{1}\bar{2}}$, scaled by $F_{12}^{T=0}=-23\alpha_1\alpha_2/64\pi^3a^7$, satisfy
\begin{equation}
\label{eqRD.p3.9}
F_{1\bar{2}}^{T=0}+F_{\bar{1}2}^{T=0}=-\frac{64(1+4r)}{r^3(1+r)^4}
,\ F_{\bar{1}\bar{2}}^{T=0}=\frac{1}{r^7},
\end{equation}
with $r=\sqrt{1+4Z^2/a^2}$, $Z$ and $a$ being the particle-mirror and particle-particle distances respectively. Eq.~\eqref{eqRD.p3.9} gives exactly Eq.~(21) in Ref.~[\onlinecite{milton2013three}]. The low-temperature corrections are
\begin{equation}
\label{eqRD.p3.10}
F_{1\bar{2}}^{T}+F_{\bar{1}2}^{T}=\frac{2r^{2}-3-r^{5}}{45r^5}(2\pi aT)^4
,\ F_{\bar{1}\bar{2}}^{T}=\frac{11(2\pi aT)^6}{945r}.
\end{equation}
Therefore, quantities in Eq.~\eqref{eqRD.p3.9} are actually 2-path contributions from anomalous paths induced by the mirror, even though three bodies are playing their own roles in those paths~\cite{milton2013three}.

\par The thermal correction to the Casimir force induced by multi-particle interactions can be evaluated in terms of the Casimir entropy, as with the two-particle cases above. For brevity, we here will not go in depth into this aspect, which could have considerable impacts in various applications.

\par As shown above, three main factors, namely the temperature, the length of the path and the orientation of the path, determine the properties of thermal Casimir interactions. The length of path, together with the temperature, sets the scale of the thermal Casimir interaction, while the orientation of path vitally influences its properties, such as its negativity. Even with this simple model for attributing contributions, however, for paths with geometry more complex than those in the simple cases treated above, the analysis is still too complicated to be claimed to be completely under control. It is worth mentioning that the multi-scattering picture gives us a conceptual interpretation of the nonadditivity of Casimir interactions, which is a long recognized phenomenon, and results in many complexities. Adding a particle to or subtracting it from a multi-particle system will create or destroy various scattering paths, and contributions from those paths, especially $n$-paths ($n\geq3$), cause significant deviations from  traditional pairwise additivity.

\par Nonetheless, there are still some reservations as to the practical significance of multi-particle Casimir interactions. It can be well-argued that the multi-particle contributions are typically too small to be detected experimentally to date, in systems comprising of a few nanoparticles, as some previous works~\cite{milton2013three,milton2015three} have claimed. Moreover, it is well-known, and recently justified again~\cite{milton2020self}, that even in some continuum scenarios, where the media involved are dilute, it is good enough to approximate the Casimir interactions with pairwise summation, by ignoring multi-particle contributions. Yet the multi-particle results, complicating the Casimir interactions inside and among continua, are usually far from trivial. Moreover, their thermal corrections should be better understood. The study of the three-particle system above is just a preliminary attempt in this direction, and more endeavors are to be made. As one small step forward, in the following Section we apply our scheme to investigate a model with more particles. There, we show that the thermal correction to multi-particle free energies can be quite important.

\subsection{Particle near a Dirac lattice}
\label{RD.psl}
\begin{figure}
  \centering
  \includegraphics[scale=0.33]{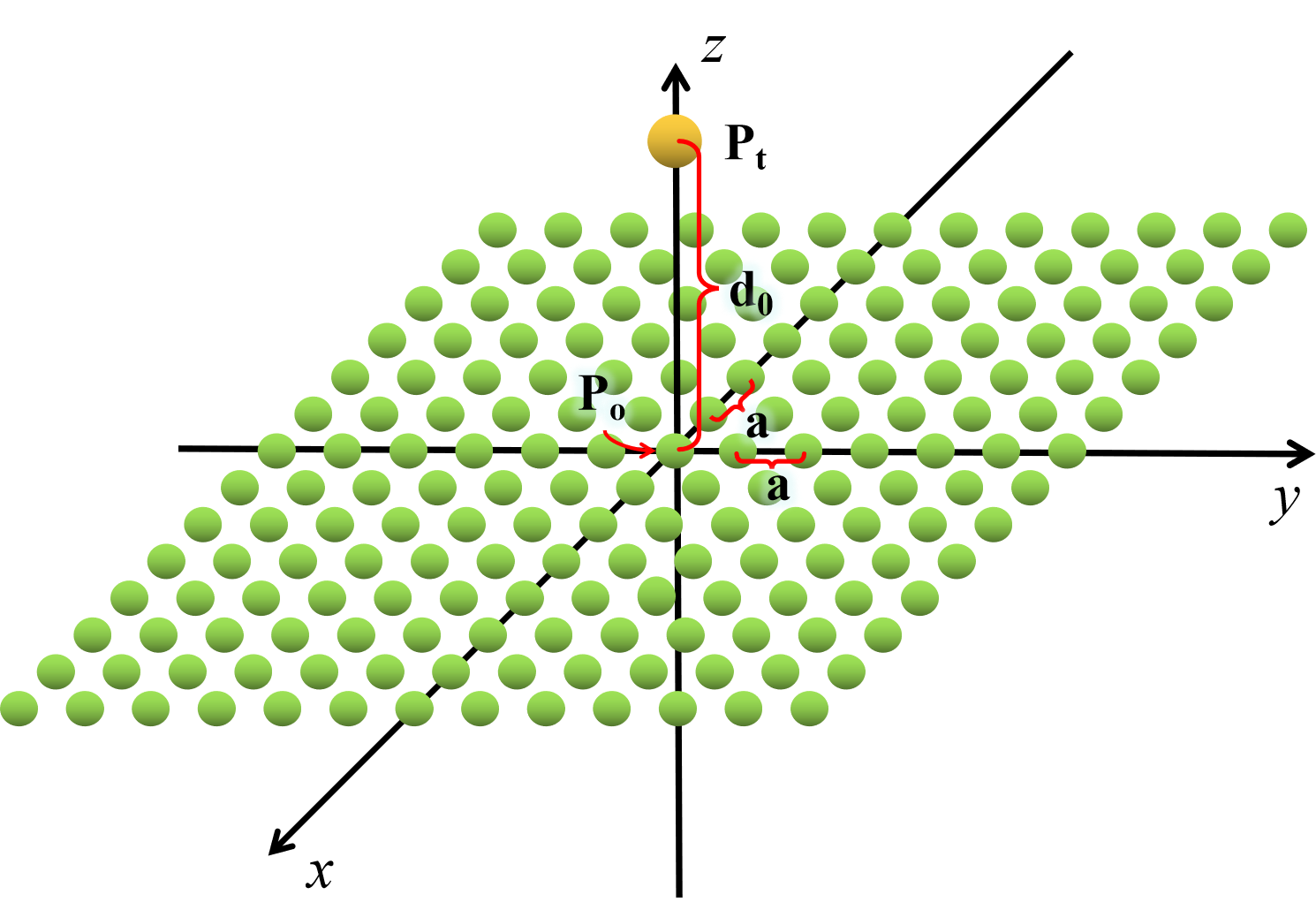}
  \caption{\label{PlPt}An isotropic nondispersive particle $\rm{P}_t$ located at $d_0\hat{\mathbf{z}}$ with the radius $\rm 25\ nm$ and polarizability $\alpha$ satisfying $\alpha/\varepsilon_0=100000\rm\ nm^3$, in front of a $13\times13$ square lattice in the $xy$-plane with the lattice vectors $\mathbf{a}_1=a\hat{\mathbf{x}},\ \mathbf{a}_2=a\hat{\mathbf{y}},\ a=250\rm\ nm$. There is one nanoparticle sitting at each lattice point.}
\end{figure}
\par Since nanoparticles have a wide range of applications, various studies on nanoparticles have been carried out. For instance, efforts have been devoted to schemes for determining the polarizability of nanoparticles recently \cite{cao2019measuring,mader2022quantitative}. Configurations built from nanoparticles are also gaining more and more attention. Lattices of nanoparticles, as an example, are modeled with properly arranged $\delta$-potentials (Dirac lattices) and the Casimir interactions between them at zero temperature have been
explored~\cite{bordag2017casimir,pirozhenko2020finite}. With the aim of illustrating the non-triviality of multi-particle contributions, particularly at finite temperature, here we numerically investigate the Casimir interaction between an isotropic nondispersive nanoparticle (labeled as $\rm{P}_t$) and a small 2D lattice with one nonspersive
nanoparticle sitting on each lattice point, as schematically illustrated  in Fig.~\ref{PlPt}. We consider nanoparticles in this lattice  capable of
being polarized either isotropically (isotropic lattice, IL) or isotropically only in the plane  of the lattice (transverse lattice, TL). To theoretically evaluate the Casimir interactions in this model, we employ the radius and polarizability of the particles listed in the caption of Fig.~\ref{PlPt}, based on the scales of those parameters in
Ref.~[\onlinecite{mader2022quantitative}].
\begin{figure*}
  \centering
  \includegraphics[scale=1.55]{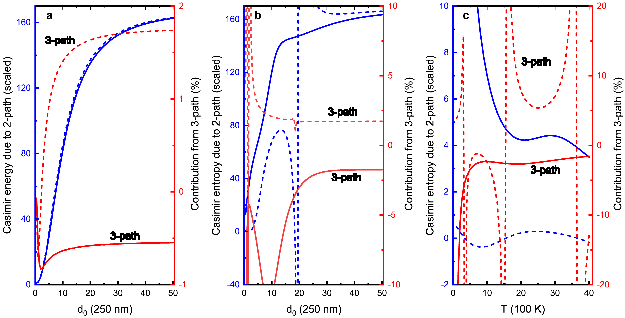}
  \caption{\label{FETZDep}For the model in Fig.~\ref{PlPt}, let $\rm{P}_t$ be located at $d_0\hat{\mathbf{z}}$ with $d_0$ being the distance between $\rm{P}_t$ and the particle at the point of origin, $\rm{P}_o$, and consider two types of lattices, namely the isotropic lattice (IL, solid lines) and the transverse lattice (TL, dashed lines). Lattice particles of IL and TL have same settings as $\rm{P}_t$, but those of TL can not polarize perpendicularly to the lattice. The contributions from 2-paths (blue) and 3-paths (red, also labeled) to quantities are evaluated. 2-path results are scaled by their own $\rm{P}_t-\rm{P}_o$ 2-path contributions, while 3-path results are shown as ratios to their corresponding 2-path results. \textbf{(a)}\ The Casimir energies at zero temperature as functions of the distance between $\rm{P}_t$ and the lattice $d_0$. \textbf{(b)}\ Casimir entropies at $T=300\rm\ K$ as functions of $d_0$. \textbf{(c)} Casimir entropies when $d_0=a=250\rm\ nm$ as functions of temperature.}
\end{figure*}

\par Suppose $\rm{P}_t$ is always on the positive $z$-axis, then we show the dependence of Casimir energy and entropy on the separation between $\rm{P}_t$ and the lattice in Fig.~\ref{FETZDep}. The 2-path Casimir energy presented in the scaled forms in Fig.~\ref{FETZDep}a, for either the $\rm{P}_t$-IL or $\rm{P}_t$-TL case, starts from $1$ and approaches the particle number of the lattice, $169$. So, if $\rm{P}_t$ is close to the lattice, the result tends to two-particle Casimir-Polder energy between $\rm{P}_t$ and the particle at the center of the lattice $\rm{P}_o$, agreeing with the results in Ref.~[\onlinecite{bordag2017casimir}], and the sensitive influence of the length of path on Casimir interaction is evident. When $\rm{P}_t$ is far away from the lattice, the lattice acts as a single particle with the polarizablity $169$ times that of a single lattice particle. 3-path contributions at zero temperature are typically small, that is, about $1\%$ of their 2-path counterparts, as predicted by previous works. But this does not hold true for thermal Casimir interactions, as revealed by the Casimir entropy, shown in Fig.~\ref{FETZDep}b and c. Similar small- and large-separation behaviors to Fig.~\ref{FETZDep}a are seen in 2-path Casimir entropy for both $\rm{P}_t$-IL and $\rm{P}_t$-TL. The scaled 2-path Casimir entropy of $\rm{P}_t$-IL converges to one more slowly than that of $\rm{P}_t$-TL. This $\rm{P}_t$-TL entropy does not change sign, although the $\rm{P}_t$-$\rm{P}_o$ two-particle entropy passes through zero at about $d_0\approx5\rm\ \mu m$ at $T=300\rm\ K$, as shown by $s_{\perp}$ in Fig.~\ref{figD01}. Furthermore, Fig.~\ref{FETZDep}b also unveils the
remarkable significance of the 3-path thermal Casimir interactions, in contrast with the zero-temperature cases. The singular behaviors of the relative 3-path entropies for the $\rm{P}_t$-TL case are due to the vanishing of the corresponding 2-path contributions. If the separation between $\rm{P}_t$ and the lattice is fixed and the temperature is varied, as shown in Fig.~\ref{FETZDep}c, the nonmonotonicity of 2-path entropies is clear, implying the subtlety of thermal Casimir interactions; for instance, their contributions to the specific heat of the system can be negative. The dependence on temperature of 3-path interactions is nontrivial also, such as its changing from negative to positive in the $\rm{P}_t$-IL case. Therefore, when evaluating the thermal Casimir interactions in a multi-particle system, contributions from $n$-paths should be properly taken into account. The combined influences of the geometry of the path and the polarizability of the particle should be carefully assessed.

\par Assume now that the distance between $\rm{P}_t$ and the lattice is fixed, at $d_0=250\rm\ nm$ for instance, and the position of $\rm{P}_t$ can vary horizontally.
At zero temperature, the results for $\rm{P}_t$-IL and $\rm{P}_t$-TL, in Fig.~\ref{HDep}a, b, c, and d, are qualitatively analogous to each other, since $d_0$ is relatively small. The total Casimir energy, from both 2-path and 3-path scatterings, demonstrates the periodicity of the lattice as the position of $\rm{P}_t$ changes, and the 3-path Casimir energies are typically orders of magnitude smaller than their corresponding 2-path counterparts. At finite temperature, on the contrary, this periodicity tends to disappear, as shown in Fig.~\ref{HDep}e, f, g and h. The 2-path Casimir entropy of $\rm{P}_t$-IL does not change sign and decreases from the center of lattice towards the outside, while that of $\rm{P}_t$-TL can be either negative or positive, depending on where $\rm{P}_t$ is with respect to this small lattice. The 3-path Casimir entropies, in Fig.~\ref{HDep}f and h, are too large to be ignored. The regions in Fig.~\ref{HDep}h where the entropy is out of  scale, results from the zeros in Fig.~\ref{HDep}g.
\begin{figure*}
  \centering
  \includegraphics[scale=0.35]{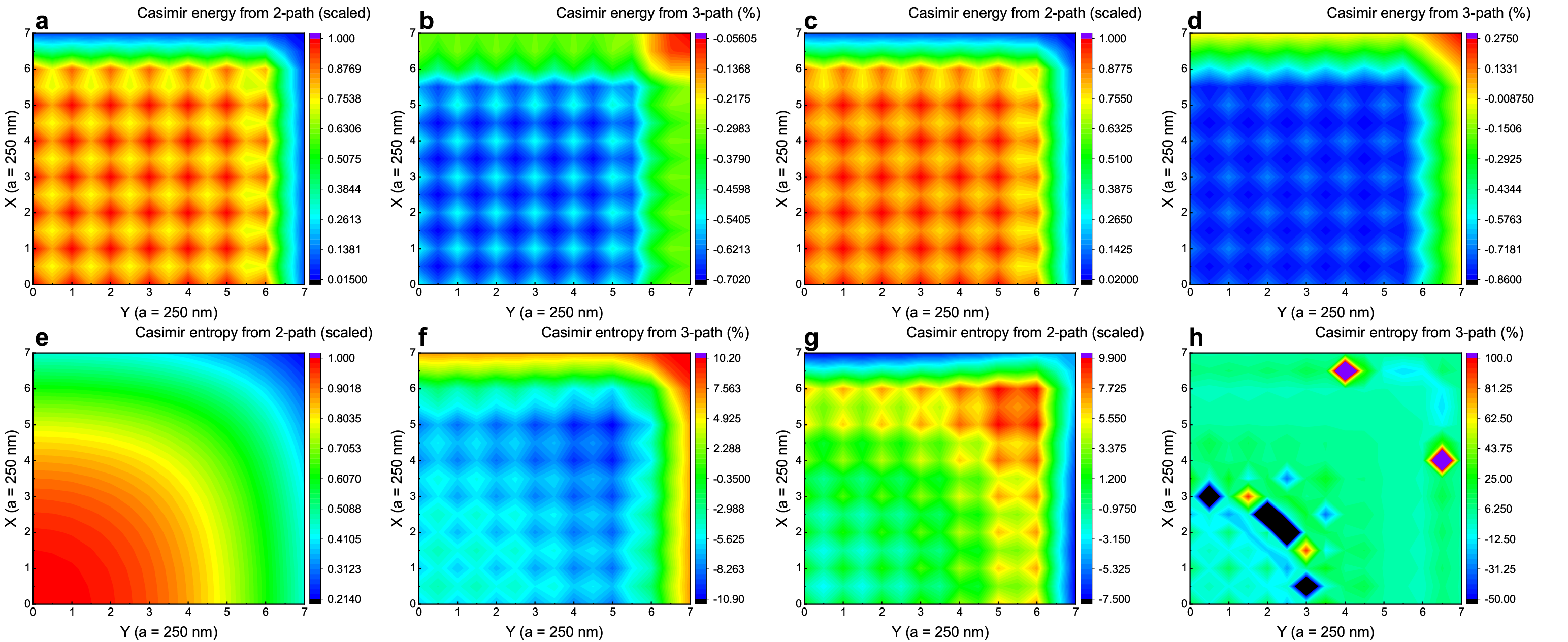}
  \caption{\label{HDep}For the model in Fig.~\ref{PlPt},  the distance between $\rm{P}_t$ and the lattice is fixed to be $a=250\rm nm$, and  the horizontal position of $\rm{P}_t$ in the first quadrant of $xy$-plane is varied. Two types of lattices as in Fig.~\ref{FETZDep} are used, namely the isotropic lattice (IL, \textbf{a}, \textbf{b}, \textbf{e}, \textbf{f}) and the transverse lattice (TL, \textbf{c}, \textbf{d}, \textbf{g}, \textbf{h}). 3-path results are exhibited as ratios in the same manner as in Fig.~\ref{FETZDep}, while 2-path results are scaled by their values when $\rm{P}_t$ is on the $z$-axis. \textbf{a}, \textbf{b}, \textbf{c}, \textbf{d} show the Casimir energies at zero temperature as functions of the horizontal position of $\rm{P}_t$, while \textbf{e}, \textbf{f}, \textbf{g}, \textbf{h} show the Casimir entropies at $T=300\rm\ K$ as functions of the horizontal position of $\rm{P}_t$.}
\end{figure*}

\par Within this simple model, the significance of $n$-path contributions to thermal Casimir interactions are apparent, even though their zero-temperature counterparts are rather small. In practical application scenarios, where the thermal Casimir interactions is important, multi-path contributions should be analyzed and handled cautiously.

\subsection{Limit to continuum}
\label{RD.DSB}
\par One may expect to model a contiuous medium by increasing the number of nanoparticles in the lattice. In a system made up of multitudinous polarizable particles, the effects of multi-particle interactions are nontrivial, if not dominating. To give a preliminary example in this direction at the end of this current work, we discuss briefly a well-established medium, which is a bulk medium characterized by the isotropic, nondispersive permittivity $\varepsilon$ and permeability $\mu=1$. Model this continuum with a cubic Dirac lattice with the basis vectors $\mathbf{a}_1=a\hat{\mathbf{x}},\ \mathbf{a}_2=a\hat{\mathbf{y}},\ \mathbf{a}_3=a\hat{\mathbf{z}}$, and the susceptibility of each particle $\bm{\alpha}=\alpha\bm{1}$ satisfying (in the dilute appproximation)
\begin{equation}
\label{eqRD.DSB.1}
\varepsilon-1=\frac{\alpha}{a^3},\quad \alpha\ll a^3,
\end{equation}
implying the convergence to the localized continuum as $a\rightarrow0$, that is, $a$ is the smallest scale, \textcolor{red}{} apart from $\alpha$ which always must be smaller than
$a^3$.}
When the temperature is zero, the magnitude of Casimir energy  (energy density) from each channel is proportional to $a^{-1}$ ($a^{-4}$).
For instance, the contribution from a channel of a 3-path as in Eq.~\eqref{eqRD.p3.6} is now
\begin{eqnarray}
\label{eqRD.DSB.2}
U_{\mathfrak{P}}^{\mu\nu\lambda}
=\frac{C_{\mathfrak{P},0}^{\mu\nu\lambda}}{a},
\end{eqnarray}
where the factor $C_{\mathfrak{P},0}^{\mu\nu\lambda}$ is, with the given $\varepsilon$, only determined by the geometry of the path, and is independent of the scaling of $a$. To show this clearly, $C_{\mathfrak{P},0}^{\mu\nu\lambda}$ can be explicitly written as
\begin{eqnarray}
\label{eqRD.DSB.3}
C_{\mathfrak{P},0}^{\mu\nu\lambda}
&=&
\frac{(\varepsilon-1)^3}{128\pi^4\rho^{10}}\frac{\gamma_{\mu\nu\lambda}
}{\eta_1^3\eta_2^3\eta_3^3}
\int^{\infty}_0dx\frac{y_{\mu}(\eta_1x)
y_{\nu}(\eta_2x)y_{\lambda}(\eta_3x)}{e^{x}}
\nonumber\\
&=&
\frac{(\varepsilon-1)^3}{128\pi^4\rho^{10}}\frac{\gamma_{\mu\nu\lambda}
}{\eta_1^3\eta_2^3\eta_3^3}
\sum^{6}_{k=0}c_{\mathfrak{P},k}^{\mu\nu\lambda}(\bm{\eta})k!
,
\end{eqnarray}
in which $\rho=d/a$ is the ratio between the length of the path $d$ and lattice spacing $a$, and $c_{\mathfrak{P},k}^{\mu\nu\lambda}(\bm{\eta})$ is the $k$th order geometric coefficient of the given channel in the path. Here, we regard the susceptibility $\varepsilon-1$ as fixed in the continuum limit. Generally, according to the expression in Eq.~\eqref{eqT.10}, the total Casimir energy per particle is $U=Ca^{-1}$, where $C$ is a factor independent of $a$. The Casimir energy density in this medium is thereby $u=Ca^{-4}$ as $a\rightarrow0$, which is consistent with the divergent behavior obtained with the spatial point-splitting regularization. Here the Casimir self-energy of each particle has been ignored. Except for the divergences induced by the self-interaction of each particle, which theoretically causes physical difficulties, there are also divergences resulting from the idealized description of the medium, which will have no effects in real materials. The scattering paths in the medium with some nontrivial geometric configuration, such as the dielectric ball, are constrained, compared to those in a medium filling in the whole space. The features of this geometric configuration will be embedded into the properties of Casimir interaction and its thermal correction, which we would like to cover in our future work. Needless to say, issues, like the various types of particles, the spatial arrangements, etc, lead to an abundance of properties.  

\par If the system is at finite temperature, the thermal contributions, in the cubic lattice above, depend on the lattice constant $a$ with an extra scale $T$ attached, i.e. the parameter $a_T=2\pi aT$. Analyse the three-body example as in Eq.~\eqref{eqRD.p3.6} and Eq.~\eqref{eqRD.DSB.2}, then we get a similar form for the Casimir free energy
\begin{eqnarray}
\label{eqRD.DSB.4}
F_{\mathfrak{P}}^{\mu\nu\lambda}
=\frac{C_{\mathfrak{P}}^{\mu\nu\lambda}(a_T)}{a},
\end{eqnarray}
in which the coefficient $C_{\mathfrak{P}}^{\mu\nu\lambda}(a_T)$ is
\begin{eqnarray}
\label{eqRD.DSB.4b}
C_{\mathfrak{P}}^{\mu\nu\lambda}(a_T)
&=&
\frac{(\varepsilon-1)^3}{128\pi^4\rho^{10}}\frac{\gamma_{\mu\nu\lambda}
}{\eta_1^3\eta_2^3\eta_3^3}
f_{\mu\nu\lambda}(\bm{\eta};\rho a_T)
.
\end{eqnarray}
It can be easily justified that there is a $C_{\mathfrak{P},0}^{\mu\nu\lambda}=C_{\mathfrak{P}}^{\mu\nu\lambda}(0)$ in $C_{\mathfrak{P}}^{\mu\nu\lambda}(a_T)$, representing the zero-temperature contribution. The thermal coefficient $\Delta C_{\mathfrak{P}}^{\mu\nu\lambda}(a_T)=C_{\mathfrak{P}}^{\mu\nu\lambda}(a_T)-C_{\mathfrak{P}}^{\mu\nu\lambda}(0)$ is evaluated with the Euler-Maclaurin formula as  
\begin{eqnarray}
\label{eqRD.DSB.5}
& &
\Delta C_{\mathfrak{P}}^{\mu\nu\lambda}(a_T)
=
-\frac{(\varepsilon-1)^3}{128\pi^4\rho^{10}}\frac{\gamma_{\mu\nu\lambda}
}{\eta_1^3\eta_2^3\eta_3^3}
\sum^{6}_{k=0}c_{\mathfrak{P},k}^{\mu\nu\lambda}(\bm{\eta})d_T^{k+1}
\nonumber\\
& &
\times
\sum_{m=1}^{\infty}\frac{B_{2m}}{(2m)!}\frac{d^{2m-1}}{dn^{2m-1}}n^ke^{-d_{T}n}\bigg|_{n=0}
\nonumber\\
&=&
\frac{(\varepsilon-1)^3}{128\pi^4\rho^{10}}\frac{\gamma_{\mu\nu\lambda}
}{\eta_1^3\eta_2^3\eta_3^3}
\sum^{6}_{k=0}c_{\mathfrak{P},k}^{\mu\nu\lambda}(\bm{\eta})
\nonumber\\
& &
\times\bigg\{
2\mRe\bigg[\bigg(\frac{\rho a_T}{2i\pi}\bigg)^{k+1}\psi^{(k)}\bigg(\frac{i\rho a_T}{2\pi}\bigg)\bigg]-2(k!)\bigg\}
,
\end{eqnarray}
where $B_n$ is the Bernoulli number, $d_T=\rho a_T$, and $\psi^{(k)}(x)$ is the $k$th order digamma function. This result is obtained by appropriate differentiation of the generating function of the Bernoulli numbers. At the first glance, the dependence on temperature in Eq.~\eqref{eqRD.DSB.5} is rather perplexing. The last term there ensures that $C_{\mathfrak{P}}^{\mu\nu\lambda}(a_T)$ vanishes as the temperature approaches to zero, and the third law of thermodynamics is thus not violated. There remain the $T^2$ term from $k=0,1$, which are $(\rho a_T)^2/12$ and $-(\rho a_T)^2/12$ for $k=0$ and $k=1$ respectively. Fortunately, we find that for any channel in a path, its $k=0$ and $k=1$ geometric coefficient are always equal. The $T^2$ terms, consequently, always cancel at the channel level. The lowest order of $a_T$ is the fourth, since $\Delta C_{\mathfrak{P}}^{\mu\nu\lambda}(a_T)$ is an even function of $a_T$, which holds true for any channel actually.
Therefore, in the continuum limit $a\rightarrow0$, the density of free energy is then proportional to the fourth power of temperature, as is well known and already seen in Table~\ref{tb1}, and for any order of the susceptibility $\varepsilon-1$, a contribution exists, which is also consistent with the known results. The divergences seem to be not so troublesome as in the zero-temperature cases. Definitely, much more work should be devoted to this beyond the brief discussion here, if more and deeper insights are to be gained.


\section{Conclusions}
\label{C}
\par To summarize, we employ a viewpoint based on electromagnetic scattering among particles to  evaluate the multi-particle thermal Casimir interaction within systems composed of nanoparticles. The geometry of the scattering path have been delineated theoretically, and its influences on multi-particle thermal Casimir effects are illustrated with the Casimir entropy, specifically by investigating three-particle configurations. In addition, contributions from transverse, longitudinal and the mixing scatterings have been studied. In two-particle cases, transverse and mixing $2$-paths are mostly responsible for the nonmonotonicity and negativity of the Casimir entropy, while in three-particle cases, where more propagators are involved and the geometry of the $3$-path exerts its influence, the longitudinal path can contribute to negative entropy as well. Furthermore, by utilizing our scheme, we explore the Casimir interaction, especially the thermal part, in a relatively more realistic model, i.e. a nanoparticle in front of a 2D lattice. Unlike the zero-temperature Casimir interaction, in which the multi-particle contributions are typically small compared to the two-particle one, in the thermal Casimir interaction the impacts from multi-particle scatterings can hardly be thought of as trivial, and should be seriously taken into account in relevant scenarios with  ambient temperature or higher.

\par As a trial evaluation, we also study the connection between the multi-scattering perspective and the continuum in a simple model. For the continuous homogeneous medium as the background, the features of divergence of the Casimir energy have been derived, and the consistency with the form of the blackbody free energy is observed with subtle cancellations in the level of the scattering channels. In the near future, we would like to systematically investigate the aspects of Casimir interactions in continuum from the multi-scattering point of view.

\par The results obtained in this work justify the potential of this multi-particle scattering scheme in applications with complex geometric configurations and various materials, and illustrate that there exist many possibilities and means to handle, or even to design  thermal Casimir interaction. We hope that, in the near future, our proposal might be applied to more experimental accessible studies or applications, for instance, those on the pull-in instability of the micro-electromechanical or nano-electromechanical systems~\cite{sircar2020casimir}, with further-developed analytical and numerical evaluation techniques.

\begin{acknowledgments}
\par  We thank the Terahertz Physics and Devices Group, Nanchang University, for the strong computational facility support. The work of KAM was supported in part by a grant from the US National Science Foundation, No. 2008417. We thanks K. V. Shajesh for warmly inspiring discussions.
\end{acknowledgments}

\bibliography{ref_main}

\begin{thebibliography}{48}%
\makeatletter
\providecommand \@ifxundefined [1]{%
 \@ifx{#1\undefined}
}%
\providecommand \@ifnum [1]{%
 \ifnum #1\expandafter \@firstoftwo
 \else \expandafter \@secondoftwo
 \fi
}%
\providecommand \@ifx [1]{%
 \ifx #1\expandafter \@firstoftwo
 \else \expandafter \@secondoftwo
 \fi
}%
\providecommand \natexlab [1]{#1}%
\providecommand \enquote  [1]{``#1''}%
\providecommand \bibnamefont  [1]{#1}%
\providecommand \bibfnamefont [1]{#1}%
\providecommand \citenamefont [1]{#1}%
\providecommand \href@noop [0]{\@secondoftwo}%
\providecommand \href [0]{\begingroup \@sanitize@url \@href}%
\providecommand \@href[1]{\@@startlink{#1}\@@href}%
\providecommand \@@href[1]{\endgroup#1\@@endlink}%
\providecommand \@sanitize@url [0]{\catcode `\\12\catcode `\$12\catcode
  `\&12\catcode `\#12\catcode `\^12\catcode `\_12\catcode `\%12\relax}%
\providecommand \@@startlink[1]{}%
\providecommand \@@endlink[0]{}%
\providecommand \url  [0]{\begingroup\@sanitize@url \@url }%
\providecommand \@url [1]{\endgroup\@href {#1}{\urlprefix }}%
\providecommand \urlprefix  [0]{URL }%
\providecommand \Eprint [0]{\href }%
\providecommand \doibase [0]{http://dx.doi.org/}%
\providecommand \selectlanguage [0]{\@gobble}%
\providecommand \bibinfo  [0]{\@secondoftwo}%
\providecommand \bibfield  [0]{\@secondoftwo}%
\providecommand \translation [1]{[#1]}%
\providecommand \BibitemOpen [0]{}%
\providecommand \bibitemStop [0]{}%
\providecommand \bibitemNoStop [0]{.\EOS\space}%
\providecommand \EOS [0]{\spacefactor3000\relax}%
\providecommand \BibitemShut  [1]{\csname bibitem#1\endcsname}%
\let\auto@bib@innerbib\@empty
\bibitem [{\citenamefont {Casimir}(1948)}]{casimir1948attraction}%
  \BibitemOpen
  \bibfield  {author} {\bibinfo {author} {\bibfnamefont {H.~B.~G.}\
  \bibnamefont {Casimir}},\ }\bibfield  {title} {\enquote {\bibinfo {title}
  {{On the attraction between two perfectly conducting plates}},}\ }\href@noop
  {} {\bibfield  {journal} {\bibinfo  {journal} {Proc. Kon. Ned. Akad. Wet.}\
  }\textbf {\bibinfo {volume} {51}},\ \bibinfo {pages} {793} (\bibinfo {year}
  {1948})}\BibitemShut {NoStop}%
\bibitem [{\citenamefont {Milton}(2003)}]{milton2003casimir}%
  \BibitemOpen
  \bibfield  {author} {\bibinfo {author} {\bibfnamefont {K.~A.}\ \bibnamefont
  {Milton}},\ }\href@noop {} {\enquote {\bibinfo {title} {{The Casimir effect:
  physical manifestations of zero-point energy}},}\ } (\bibinfo {year}
  {2003})\BibitemShut {NoStop}%
\bibitem [{\citenamefont {Bordag}\ \emph {et~al.}(2009)\citenamefont {Bordag},
  \citenamefont {Klimchitskaya}, \citenamefont {Mohideen},\ and\ \citenamefont
  {Mostepanenko}}]{bordag2009advances}%
  \BibitemOpen
  \bibfield  {author} {\bibinfo {author} {\bibfnamefont {M.}~\bibnamefont
  {Bordag}}, \bibinfo {author} {\bibfnamefont {G.~L.}\ \bibnamefont
  {Klimchitskaya}}, \bibinfo {author} {\bibfnamefont {U.}~\bibnamefont
  {Mohideen}}, \ and\ \bibinfo {author} {\bibfnamefont {V.~M.}\ \bibnamefont
  {Mostepanenko}},\ }\href@noop {} {\emph {\bibinfo {title} {{Advances in the
  Casimir effect}}}},\ Vol.\ \bibinfo {volume} {145}\ (\bibinfo  {publisher}
  {OUP Oxford},\ \bibinfo {year} {2009})\BibitemShut {NoStop}%
\bibitem [{\citenamefont {Dalvit}\ \emph {et~al.}(2011)\citenamefont {Dalvit},
  \citenamefont {Milonni}, \citenamefont {Roberts},\ and\ \citenamefont
  {da~Rosa}}]{Dalvit2011Casimir}%
  \BibitemOpen
  \bibfield  {author} {\bibinfo {author} {\bibfnamefont {D.~A.~R.}\
  \bibnamefont {Dalvit}}, \bibinfo {author} {\bibfnamefont {P.}~\bibnamefont
  {Milonni}}, \bibinfo {author} {\bibfnamefont {R.}~\bibnamefont {Roberts}}, \
  and\ \bibinfo {author} {\bibfnamefont {F.}~\bibnamefont {da~Rosa}},\
  }\href@noop {} {\emph {\bibinfo {title} {Casimir Physics}}}\ (\bibinfo
  {publisher} {Springer, Berlin},\ \bibinfo {year} {2011})\BibitemShut
  {NoStop}%
\bibitem [{\citenamefont {Andrews}\ and\ \citenamefont
  {Bradshaw}(2014)}]{andrews2014role}%
  \BibitemOpen
  \bibfield  {author} {\bibinfo {author} {\bibfnamefont {D.~L.}\ \bibnamefont
  {Andrews}}\ and\ \bibinfo {author} {\bibfnamefont {D.~S.}\ \bibnamefont
  {Bradshaw}},\ }\bibfield  {title} {\enquote {\bibinfo {title} {The role of
  virtual photons in nanoscale photonics},}\ }\href@noop {} {\bibfield
  {journal} {\bibinfo  {journal} {Ann. Phys. (Berl.)}\ }\textbf {\bibinfo
  {volume} {526}},\ \bibinfo {pages} {173--186} (\bibinfo {year}
  {2014})}\BibitemShut {NoStop}%
\bibitem [{\citenamefont {Rodriguez}\ \emph {et~al.}(2015)\citenamefont
  {Rodriguez}, \citenamefont {Hui}, \citenamefont {Woolf}, \citenamefont
  {Johnson}, \citenamefont {Lon{\v{c}}ar},\ and\ \citenamefont
  {Capasso}}]{rodriguez2015classical}%
  \BibitemOpen
  \bibfield  {author} {\bibinfo {author} {\bibfnamefont {Alejandro~W}\
  \bibnamefont {Rodriguez}}, \bibinfo {author} {\bibfnamefont {Pui-Chuen}\
  \bibnamefont {Hui}}, \bibinfo {author} {\bibfnamefont {David~P}\ \bibnamefont
  {Woolf}}, \bibinfo {author} {\bibfnamefont {Steven~G}\ \bibnamefont
  {Johnson}}, \bibinfo {author} {\bibfnamefont {Marko}\ \bibnamefont
  {Lon{\v{c}}ar}}, \ and\ \bibinfo {author} {\bibfnamefont {Federico}\
  \bibnamefont {Capasso}},\ }\bibfield  {title} {\enquote {\bibinfo {title}
  {{Classical and fluctuation-induced electromagnetic interactions in
  micron-scale systems: designer bonding, antibonding, and Casimir forces}},}\
  }\href@noop {} {\bibfield  {journal} {\bibinfo  {journal} {Ann. Phys.
  (Berl.)}\ }\textbf {\bibinfo {volume} {527}},\ \bibinfo {pages} {45--80}
  (\bibinfo {year} {2015})}\BibitemShut {NoStop}%
\bibitem [{\citenamefont {Esteso}\ \emph {et~al.}(2020)\citenamefont {Esteso},
  \citenamefont {Carretero-Palacios}, \citenamefont {MacDowell}, \citenamefont
  {Fiedler}, \citenamefont {Parsons}, \citenamefont {Spallek}, \citenamefont
  {M{\'\i}guez}, \citenamefont {Persson}, \citenamefont {Buhmann},
  \citenamefont {Brevik},\ and\ \citenamefont
  {Bostr\"{o}m}}]{esteso2020premelting}%
  \BibitemOpen
  \bibfield  {author} {\bibinfo {author} {\bibfnamefont {V.}~\bibnamefont
  {Esteso}}, \bibinfo {author} {\bibfnamefont {S.}~\bibnamefont
  {Carretero-Palacios}}, \bibinfo {author} {\bibfnamefont {L.~G.}\ \bibnamefont
  {MacDowell}}, \bibinfo {author} {\bibfnamefont {J.}~\bibnamefont {Fiedler}},
  \bibinfo {author} {\bibfnamefont {D.~F.}\ \bibnamefont {Parsons}}, \bibinfo
  {author} {\bibfnamefont {F.}~\bibnamefont {Spallek}}, \bibinfo {author}
  {\bibfnamefont {H.}~\bibnamefont {M{\'\i}guez}}, \bibinfo {author}
  {\bibfnamefont {C.}~\bibnamefont {Persson}}, \bibinfo {author} {\bibfnamefont
  {S.~Y.}\ \bibnamefont {Buhmann}}, \bibinfo {author} {\bibfnamefont
  {I.}~\bibnamefont {Brevik}}, \ and\ \bibinfo {author} {\bibfnamefont
  {M.}~\bibnamefont {Bostr\"{o}m}},\ }\bibfield  {title} {\enquote {\bibinfo
  {title} {Premelting of ice adsorbed on a rock surface},}\ }\href@noop {}
  {\bibfield  {journal} {\bibinfo  {journal} {Phys. Chem. Chem. Phys.}\
  }\textbf {\bibinfo {volume} {22}},\ \bibinfo {pages} {11362--11373} (\bibinfo
  {year} {2020})}\BibitemShut {NoStop}%
\bibitem [{\citenamefont {Luengo-M{\'a}rquez}\ and\ \citenamefont
  {MacDowell}(2021)}]{luengo2021lifshitz}%
  \BibitemOpen
  \bibfield  {author} {\bibinfo {author} {\bibfnamefont {J.}~\bibnamefont
  {Luengo-M{\'a}rquez}}\ and\ \bibinfo {author} {\bibfnamefont {L.~G.}\
  \bibnamefont {MacDowell}},\ }\bibfield  {title} {\enquote {\bibinfo {title}
  {{Lifshitz theory of wetting films at three phase coexistence: The case of
  ice nucleation on Silver Iodide (AgI)}},}\ }\href@noop {} {\bibfield
  {journal} {\bibinfo  {journal} {J. Colloid and Interface Sci.}\ }\textbf
  {\bibinfo {volume} {590}},\ \bibinfo {pages} {527--538} (\bibinfo {year}
  {2021})}\BibitemShut {NoStop}%
\bibitem [{\citenamefont {Zhao}\ \emph {et~al.}(2019)\citenamefont {Zhao},
  \citenamefont {Li}, \citenamefont {Yang}, \citenamefont {Bao}, \citenamefont
  {Xia}, \citenamefont {Ashby}, \citenamefont {Wang},\ and\ \citenamefont
  {Zhang}}]{zhao2019stable}%
  \BibitemOpen
  \bibfield  {author} {\bibinfo {author} {\bibfnamefont {R.}~\bibnamefont
  {Zhao}}, \bibinfo {author} {\bibfnamefont {L.}~\bibnamefont {Li}}, \bibinfo
  {author} {\bibfnamefont {S.}~\bibnamefont {Yang}}, \bibinfo {author}
  {\bibfnamefont {W.}~\bibnamefont {Bao}}, \bibinfo {author} {\bibfnamefont
  {Y.}~\bibnamefont {Xia}}, \bibinfo {author} {\bibfnamefont {P.}~\bibnamefont
  {Ashby}}, \bibinfo {author} {\bibfnamefont {Y.}~\bibnamefont {Wang}}, \ and\
  \bibinfo {author} {\bibfnamefont {X.}~\bibnamefont {Zhang}},\ }\bibfield
  {title} {\enquote {\bibinfo {title} {{Stable Casimir equilibria and quantum
  trapping}},}\ }\href@noop {} {\bibfield  {journal} {\bibinfo  {journal}
  {Science}\ }\textbf {\bibinfo {volume} {364}},\ \bibinfo {pages} {984--987}
  (\bibinfo {year} {2019})}\BibitemShut {NoStop}%
\bibitem [{\citenamefont {M.}(2011)}]{Schaden2011Irreducible}%
  \BibitemOpen
  \bibfield  {author} {\bibinfo {author} {\bibfnamefont {Schaden}\ \bibnamefont
  {M.}},\ }\bibfield  {title} {\enquote {\bibinfo {title} {Irreducible
  many-body {Casimir} energies of intersecting objects},}\ }\href@noop {}
  {\bibfield  {journal} {\bibinfo  {journal} {EPL}\ }\textbf {\bibinfo {volume}
  {94}},\ \bibinfo {pages} {41001} (\bibinfo {year} {2011})}\BibitemShut
  {NoStop}%
\bibitem [{\citenamefont {Shajesh}\ and\ \citenamefont
  {Schaden}(2011)}]{Shajesh2011Many}%
  \BibitemOpen
  \bibfield  {author} {\bibinfo {author} {\bibfnamefont {K.~V.}\ \bibnamefont
  {Shajesh}}\ and\ \bibinfo {author} {\bibfnamefont {M.}~\bibnamefont
  {Schaden}},\ }\bibfield  {title} {\enquote {\bibinfo {title} {{Many-body
  contributions to Green's functions and Casimir energies}},}\ }\href@noop {}
  {\bibfield  {journal} {\bibinfo  {journal} {Phys. Rev. D}\ }\textbf {\bibinfo
  {volume} {83}},\ \bibinfo {pages} {527--542} (\bibinfo {year}
  {2011})}\BibitemShut {NoStop}%
\bibitem [{\citenamefont {Rodriguez-Lopez}\ \emph {et~al.}(2009)\citenamefont
  {Rodriguez-Lopez}, \citenamefont {Rahi},\ and\ \citenamefont
  {Emig}}]{rodriguez2009three}%
  \BibitemOpen
  \bibfield  {author} {\bibinfo {author} {\bibfnamefont {P.}~\bibnamefont
  {Rodriguez-Lopez}}, \bibinfo {author} {\bibfnamefont {S.~J.}\ \bibnamefont
  {Rahi}}, \ and\ \bibinfo {author} {\bibfnamefont {T.}~\bibnamefont {Emig}},\
  }\bibfield  {title} {\enquote {\bibinfo {title} {{Three-body Casimir effects
  and nonmonotonic forces}},}\ }\href@noop {} {\bibfield  {journal} {\bibinfo
  {journal} {Phys. Rev. A}\ }\textbf {\bibinfo {volume} {80}},\ \bibinfo
  {pages} {022519} (\bibinfo {year} {2009})}\BibitemShut {NoStop}%
\bibitem [{\citenamefont {Shajesh}\ and\ \citenamefont
  {Schaden}(2012)}]{shajesh2012significance}%
  \BibitemOpen
  \bibfield  {author} {\bibinfo {author} {\bibfnamefont {K.~V.}\ \bibnamefont
  {Shajesh}}\ and\ \bibinfo {author} {\bibfnamefont {M.}~\bibnamefont
  {Schaden}},\ }\bibfield  {title} {\enquote {\bibinfo {title} {{Significance
  of many-body contributions to Casimir energies}},}\ }in\ \href@noop {} {\emph
  {\bibinfo {booktitle} {Int. J. Mod. Phys. Conf. Ser.}}},\ Vol.~\bibinfo
  {volume} {14}\ (\bibinfo {organization} {World Scientific},\ \bibinfo {year}
  {2012})\ pp.\ \bibinfo {pages} {521--530}\BibitemShut {NoStop}%
\bibitem [{\citenamefont {Milton}\ \emph
  {et~al.}(2015{\natexlab{a}})\citenamefont {Milton}, \citenamefont {Abalo},
  \citenamefont {Parashar}, \citenamefont {Pourtolami},\ and\ \citenamefont
  {Scheel}}]{Milton2015Casimir}%
  \BibitemOpen
  \bibfield  {author} {\bibinfo {author} {\bibfnamefont {K.~A.}\ \bibnamefont
  {Milton}}, \bibinfo {author} {\bibfnamefont {E.~K.}\ \bibnamefont {Abalo}},
  \bibinfo {author} {\bibfnamefont {P.}~\bibnamefont {Parashar}}, \bibinfo
  {author} {\bibfnamefont {N.}~\bibnamefont {Pourtolami}}, \ and\ \bibinfo
  {author} {\bibfnamefont {S.}~\bibnamefont {Scheel}},\ }\bibfield  {title}
  {\enquote {\bibinfo {title} {Casimir-polder repulsion: Three-body effects},}\
  }\href@noop {} {\bibfield  {journal} {\bibinfo  {journal} {Phys. Rev. A}\
  }\textbf {\bibinfo {volume} {91}},\ \bibinfo {pages} {042510} (\bibinfo
  {year} {2015}{\natexlab{a}})}\BibitemShut {NoStop}%
\bibitem [{\citenamefont {Xu}\ \emph {et~al.}(2022)\citenamefont {Xu},
  \citenamefont {Ju}, \citenamefont {Gao}, \citenamefont {Shen}, \citenamefont
  {Jacob},\ and\ \citenamefont {Li}}]{xu2022observation}%
  \BibitemOpen
  \bibfield  {author} {\bibinfo {author} {\bibfnamefont {Z.}~\bibnamefont
  {Xu}}, \bibinfo {author} {\bibfnamefont {P.}~\bibnamefont {Ju}}, \bibinfo
  {author} {\bibfnamefont {X.}~\bibnamefont {Gao}}, \bibinfo {author}
  {\bibfnamefont {K.}~\bibnamefont {Shen}}, \bibinfo {author} {\bibfnamefont
  {Z.}~\bibnamefont {Jacob}}, \ and\ \bibinfo {author} {\bibfnamefont
  {T.}~\bibnamefont {Li}},\ }\bibfield  {title} {\enquote {\bibinfo {title}
  {{Observation and control of Casimir effects in a sphere-plate-sphere
  system}},}\ }\href@noop {} {\bibfield  {journal} {\bibinfo  {journal} {Nat.
  Commun.}\ }\textbf {\bibinfo {volume} {13}},\ \bibinfo {pages} {6148}
  (\bibinfo {year} {2022})}\BibitemShut {NoStop}%
\bibitem [{\citenamefont {Venkataram}\ \emph {et~al.}(2019)\citenamefont
  {Venkataram}, \citenamefont {Hermann}, \citenamefont {Vongkovit},
  \citenamefont {Tkatchenko},\ and\ \citenamefont
  {Rodriguez}}]{venkataram2019impact}%
  \BibitemOpen
  \bibfield  {author} {\bibinfo {author} {\bibfnamefont {P.~S.}\ \bibnamefont
  {Venkataram}}, \bibinfo {author} {\bibfnamefont {J.}~\bibnamefont {Hermann}},
  \bibinfo {author} {\bibfnamefont {T.~J.}\ \bibnamefont {Vongkovit}}, \bibinfo
  {author} {\bibfnamefont {A.}~\bibnamefont {Tkatchenko}}, \ and\ \bibinfo
  {author} {\bibfnamefont {A.~W.}\ \bibnamefont {Rodriguez}},\ }\bibfield
  {title} {\enquote {\bibinfo {title} {{Impact of nuclear vibrations on van der
  Waals and Casimir interactions at zero and finite temperature}},}\
  }\href@noop {} {\bibfield  {journal} {\bibinfo  {journal} {Science Advances}\
  }\textbf {\bibinfo {volume} {5}},\ \bibinfo {pages} {eaaw0456} (\bibinfo
  {year} {2019})}\BibitemShut {NoStop}%
\bibitem [{\citenamefont {Decca}\ \emph {et~al.}(2005)\citenamefont {Decca},
  \citenamefont {L\'{o}pez}, \citenamefont {Fischbach}, \citenamefont
  {Klimchitskaya}, \citenamefont {Krause},\ and\ \citenamefont
  {Mostepanenko}}]{Decca2005Precise}%
  \BibitemOpen
  \bibfield  {author} {\bibinfo {author} {\bibfnamefont {R.~S.}\ \bibnamefont
  {Decca}}, \bibinfo {author} {\bibfnamefont {D.}~\bibnamefont {L\'{o}pez}},
  \bibinfo {author} {\bibfnamefont {E.}~\bibnamefont {Fischbach}}, \bibinfo
  {author} {\bibfnamefont {G.~L.}\ \bibnamefont {Klimchitskaya}}, \bibinfo
  {author} {\bibfnamefont {D.~E.}\ \bibnamefont {Krause}}, \ and\ \bibinfo
  {author} {\bibfnamefont {V.~M.}\ \bibnamefont {Mostepanenko}},\ }\bibfield
  {title} {\enquote {\bibinfo {title} {{Precise comparison of theory and new
  experiment for the Casimir force leads to stronger constraints on thermal
  quantum effects and long-range interactions}},}\ }\href@noop {} {\bibfield
  {journal} {\bibinfo  {journal} {Ann. Phys. (N. Y.)}\ }\textbf {\bibinfo
  {volume} {318}},\ \bibinfo {pages} {37--80} (\bibinfo {year}
  {2005})}\BibitemShut {NoStop}%
\bibitem [{\citenamefont {Liu}\ \emph {et~al.}(2019)\citenamefont {Liu},
  \citenamefont {Xu}, \citenamefont {Klimchitskaya}, \citenamefont
  {Mostepanenko},\ and\ \citenamefont {Mohideen}}]{Liu2019Examining}%
  \BibitemOpen
  \bibfield  {author} {\bibinfo {author} {\bibfnamefont {M.}~\bibnamefont
  {Liu}}, \bibinfo {author} {\bibfnamefont {J.}~\bibnamefont {Xu}}, \bibinfo
  {author} {\bibfnamefont {G.~L.}\ \bibnamefont {Klimchitskaya}}, \bibinfo
  {author} {\bibfnamefont {V.~M.}\ \bibnamefont {Mostepanenko}}, \ and\
  \bibinfo {author} {\bibfnamefont {U.}~\bibnamefont {Mohideen}},\ }\bibfield
  {title} {\enquote {\bibinfo {title} {{Examining the Casimir puzzle with
  upgraded technique and advanced surface cleaning}},}\ }\href@noop {}
  {\bibfield  {journal} {\bibinfo  {journal} {Phys. Rev. B}\ }\textbf {\bibinfo
  {volume} {100}},\ \bibinfo {pages} {081406} (\bibinfo {year}
  {2019})}\BibitemShut {NoStop}%
\bibitem [{\citenamefont {Sushkov}\ \emph {et~al.}(2011)\citenamefont
  {Sushkov}, \citenamefont {Kim}, \citenamefont {Dalvit},\ and\ \citenamefont
  {Lamoreaux}}]{Sushkov2011Observation}%
  \BibitemOpen
  \bibfield  {author} {\bibinfo {author} {\bibfnamefont {A.~O.}\ \bibnamefont
  {Sushkov}}, \bibinfo {author} {\bibfnamefont {W.~J.}\ \bibnamefont {Kim}},
  \bibinfo {author} {\bibfnamefont {Dar}\ \bibnamefont {Dalvit}}, \ and\
  \bibinfo {author} {\bibfnamefont {S.~K.}\ \bibnamefont {Lamoreaux}},\
  }\bibfield  {title} {\enquote {\bibinfo {title} {{Observation of the thermal
  Casimir force}},}\ }\href@noop {} {\bibfield  {journal} {\bibinfo  {journal}
  {Nat. Phys.}\ }\textbf {\bibinfo {volume} {7}},\ \bibinfo {pages} {230--233}
  (\bibinfo {year} {2011})}\BibitemShut {NoStop}%
\bibitem [{\citenamefont {Garcia-Sanchez}\ \emph {et~al.}(2012)\citenamefont
  {Garcia-Sanchez}, \citenamefont {Fong}, \citenamefont {Bhaskaran},
  \citenamefont {Lamoreaux},\ and\ \citenamefont {Hong}}]{Garcia2012Casimir}%
  \BibitemOpen
  \bibfield  {author} {\bibinfo {author} {\bibfnamefont {D.}~\bibnamefont
  {Garcia-Sanchez}}, \bibinfo {author} {\bibfnamefont {K.~Y.}\ \bibnamefont
  {Fong}}, \bibinfo {author} {\bibfnamefont {H.}~\bibnamefont {Bhaskaran}},
  \bibinfo {author} {\bibfnamefont {S.}~\bibnamefont {Lamoreaux}}, \ and\
  \bibinfo {author} {\bibfnamefont {X.~T.}\ \bibnamefont {Hong}},\ }\bibfield
  {title} {\enquote {\bibinfo {title} {Casimir force and in situ surface
  potential measurements on nanomembranes.}}\ }\href@noop {} {\bibfield
  {journal} {\bibinfo  {journal} {Phys. Rev. Lett.}\ }\textbf {\bibinfo
  {volume} {109}},\ \bibinfo {pages} {027202} (\bibinfo {year}
  {2012})}\BibitemShut {NoStop}%
\bibitem [{\citenamefont {Bezerra}\ \emph {et~al.}(2004)\citenamefont
  {Bezerra}, \citenamefont {Klimchitskaya}, \citenamefont {Mostepanenko},\ and\
  \citenamefont {Romero}}]{bezerra2004violation}%
  \BibitemOpen
  \bibfield  {author} {\bibinfo {author} {\bibfnamefont {V.~B.}\ \bibnamefont
  {Bezerra}}, \bibinfo {author} {\bibfnamefont {G.~L.}\ \bibnamefont
  {Klimchitskaya}}, \bibinfo {author} {\bibfnamefont {V.~M.}\ \bibnamefont
  {Mostepanenko}}, \ and\ \bibinfo {author} {\bibfnamefont {C.}~\bibnamefont
  {Romero}},\ }\bibfield  {title} {\enquote {\bibinfo {title} {{Violation of
  the Nernst heat theorem in the theory of the thermal Casimir force between
  Drude metals}},}\ }\href@noop {} {\bibfield  {journal} {\bibinfo  {journal}
  {Phys. Rev. A}\ }\textbf {\bibinfo {volume} {69}},\ \bibinfo {pages} {022119}
  (\bibinfo {year} {2004})}\BibitemShut {NoStop}%
\bibitem [{\citenamefont {Klimchitskaya}\ and\ \citenamefont
  {Mostepanenko}(2017)}]{klimchitskaya2017low}%
  \BibitemOpen
  \bibfield  {author} {\bibinfo {author} {\bibfnamefont {G.~L.}\ \bibnamefont
  {Klimchitskaya}}\ and\ \bibinfo {author} {\bibfnamefont {V.~M.}\ \bibnamefont
  {Mostepanenko}},\ }\bibfield  {title} {\enquote {\bibinfo {title}
  {{Low-temperature behavior of the Casimir free energy and entropy of metallic
  films}},}\ }\href@noop {} {\bibfield  {journal} {\bibinfo  {journal} {Phys.
  Rev. A}\ }\textbf {\bibinfo {volume} {95}},\ \bibinfo {pages} {012130}
  (\bibinfo {year} {2017})}\BibitemShut {NoStop}%
\bibitem [{\citenamefont {Brevik}\ \emph {et~al.}(2006)\citenamefont {Brevik},
  \citenamefont {Ellingsen},\ and\ \citenamefont {Milton}}]{brevik2006thermal}%
  \BibitemOpen
  \bibfield  {author} {\bibinfo {author} {\bibfnamefont {I.}~\bibnamefont
  {Brevik}}, \bibinfo {author} {\bibfnamefont {S.~A.}\ \bibnamefont
  {Ellingsen}}, \ and\ \bibinfo {author} {\bibfnamefont {K.~A.}\ \bibnamefont
  {Milton}},\ }\bibfield  {title} {\enquote {\bibinfo {title} {{Thermal
  corrections to the Casimir effect}},}\ }\href@noop {} {\bibfield  {journal}
  {\bibinfo  {journal} {New J. Phys.}\ }\textbf {\bibinfo {volume} {8}},\
  \bibinfo {pages} {236} (\bibinfo {year} {2006})}\BibitemShut {NoStop}%
\bibitem [{\citenamefont {Milton}\ \emph {et~al.}(2012)\citenamefont {Milton},
  \citenamefont {Brevik},\ and\ \citenamefont {Ellingsen}}]{milton2012thermal}%
  \BibitemOpen
  \bibfield  {author} {\bibinfo {author} {\bibfnamefont {K.~A.}\ \bibnamefont
  {Milton}}, \bibinfo {author} {\bibfnamefont {I.}~\bibnamefont {Brevik}}, \
  and\ \bibinfo {author} {\bibfnamefont {S.~A.}\ \bibnamefont {Ellingsen}},\
  }\bibfield  {title} {\enquote {\bibinfo {title} {{Thermal issues in Casimir
  forces between conductors and semiconductors}},}\ }\href@noop {} {\bibfield
  {journal} {\bibinfo  {journal} {Phys. Scr.}\ }\textbf {\bibinfo {volume}
  {2012}},\ \bibinfo {pages} {014070} (\bibinfo {year} {2012})}\BibitemShut
  {NoStop}%
\bibitem [{\citenamefont {Bezerra}\ \emph
  {et~al.}(2002{\natexlab{a}})\citenamefont {Bezerra}, \citenamefont
  {Klimchitskaya},\ and\ \citenamefont
  {Mostepanenko}}]{bezerra2002thermodynamical}%
  \BibitemOpen
  \bibfield  {author} {\bibinfo {author} {\bibfnamefont {V.~B.}\ \bibnamefont
  {Bezerra}}, \bibinfo {author} {\bibfnamefont {G.~L.}\ \bibnamefont
  {Klimchitskaya}}, \ and\ \bibinfo {author} {\bibfnamefont {V.~M.}\
  \bibnamefont {Mostepanenko}},\ }\bibfield  {title} {\enquote {\bibinfo
  {title} {{Thermodynamical aspects of the Casimir force between real metals at
  nonzero temperature}},}\ }\href@noop {} {\bibfield  {journal} {\bibinfo
  {journal} {Phys. Rev. A}\ }\textbf {\bibinfo {volume} {65}},\ \bibinfo
  {pages} {052113} (\bibinfo {year} {2002}{\natexlab{a}})}\BibitemShut
  {NoStop}%
\bibitem [{\citenamefont {Bezerra}\ \emph
  {et~al.}(2002{\natexlab{b}})\citenamefont {Bezerra}, \citenamefont
  {Klimchitskaya},\ and\ \citenamefont
  {Mostepanenko}}]{bezerra2002correlation}%
  \BibitemOpen
  \bibfield  {author} {\bibinfo {author} {\bibfnamefont {V.~B.}\ \bibnamefont
  {Bezerra}}, \bibinfo {author} {\bibfnamefont {G.~L.}\ \bibnamefont
  {Klimchitskaya}}, \ and\ \bibinfo {author} {\bibfnamefont {V.~M.}\
  \bibnamefont {Mostepanenko}},\ }\bibfield  {title} {\enquote {\bibinfo
  {title} {Correlation of energy and free energy for the thermal {Casimir}
  force between real metals},}\ }\href@noop {} {\bibfield  {journal} {\bibinfo
  {journal} {Phys. Rev. A}\ }\textbf {\bibinfo {volume} {66}},\ \bibinfo
  {pages} {062112} (\bibinfo {year} {2002}{\natexlab{b}})}\BibitemShut
  {NoStop}%
\bibitem [{\citenamefont {Canaguier-Durand}\ \emph {et~al.}(2010)\citenamefont
  {Canaguier-Durand}, \citenamefont {Neto}, \citenamefont {Lambrecht},\ and\
  \citenamefont {Reynaud}}]{canaguier2010thermal}%
  \BibitemOpen
  \bibfield  {author} {\bibinfo {author} {\bibfnamefont {A.}~\bibnamefont
  {Canaguier-Durand}}, \bibinfo {author} {\bibfnamefont {P.~A.~M.}\
  \bibnamefont {Neto}}, \bibinfo {author} {\bibfnamefont {A.}~\bibnamefont
  {Lambrecht}}, \ and\ \bibinfo {author} {\bibfnamefont {S.}~\bibnamefont
  {Reynaud}},\ }\bibfield  {title} {\enquote {\bibinfo {title} {{Thermal
  Casimir effect for Drude metals in the plane-sphere geometry}},}\ }\href@noop
  {} {\bibfield  {journal} {\bibinfo  {journal} {Phys. Rev. A}\ }\textbf
  {\bibinfo {volume} {82}},\ \bibinfo {pages} {012511} (\bibinfo {year}
  {2010})}\BibitemShut {NoStop}%
\bibitem [{\citenamefont {Rodriguez-Lopez}(2011)}]{rodriguez2011casimir}%
  \BibitemOpen
  \bibfield  {author} {\bibinfo {author} {\bibfnamefont {P.}~\bibnamefont
  {Rodriguez-Lopez}},\ }\bibfield  {title} {\enquote {\bibinfo {title} {Casimir
  energy and entropy in the sphere-sphere geometry},}\ }\href@noop {}
  {\bibfield  {journal} {\bibinfo  {journal} {Phys. Rev. B}\ }\textbf {\bibinfo
  {volume} {84}},\ \bibinfo {pages} {075431} (\bibinfo {year}
  {2011})}\BibitemShut {NoStop}%
\bibitem [{\citenamefont {Ingold}\ \emph {et~al.}(2015)\citenamefont {Ingold},
  \citenamefont {Umrath}, \citenamefont {Hartmann}, \citenamefont
  {Gu{\'e}rout}, \citenamefont {Lambrecht}, \citenamefont {Reynaud},\ and\
  \citenamefont {Milton}}]{ingold2015geometric}%
  \BibitemOpen
  \bibfield  {author} {\bibinfo {author} {\bibfnamefont {G.}~\bibnamefont
  {Ingold}}, \bibinfo {author} {\bibfnamefont {S.}~\bibnamefont {Umrath}},
  \bibinfo {author} {\bibfnamefont {M.}~\bibnamefont {Hartmann}}, \bibinfo
  {author} {\bibfnamefont {R.}~\bibnamefont {Gu{\'e}rout}}, \bibinfo {author}
  {\bibfnamefont {A.}~\bibnamefont {Lambrecht}}, \bibinfo {author}
  {\bibfnamefont {S.}~\bibnamefont {Reynaud}}, \ and\ \bibinfo {author}
  {\bibfnamefont {Kimball~A.}\ \bibnamefont {Milton}},\ }\bibfield  {title}
  {\enquote {\bibinfo {title} {Geometric origin of negative {Casimir}
  entropies: A scattering-channel analysis},}\ }\href@noop {} {\bibfield
  {journal} {\bibinfo  {journal} {Phys. Rev. E}\ }\textbf {\bibinfo {volume}
  {91}},\ \bibinfo {pages} {033203} (\bibinfo {year} {2015})}\BibitemShut
  {NoStop}%
\bibitem [{\citenamefont {Milton}\ \emph
  {et~al.}(2015{\natexlab{b}})\citenamefont {Milton}, \citenamefont
  {Gu{\'e}rout}, \citenamefont {Ingold}, \citenamefont {Lambrecht},\ and\
  \citenamefont {Reynaud}}]{milton2015negative}%
  \BibitemOpen
  \bibfield  {author} {\bibinfo {author} {\bibfnamefont {K.~A.}\ \bibnamefont
  {Milton}}, \bibinfo {author} {\bibfnamefont {R.}~\bibnamefont {Gu{\'e}rout}},
  \bibinfo {author} {\bibfnamefont {G.}~\bibnamefont {Ingold}}, \bibinfo
  {author} {\bibfnamefont {A.}~\bibnamefont {Lambrecht}}, \ and\ \bibinfo
  {author} {\bibfnamefont {S.}~\bibnamefont {Reynaud}},\ }\bibfield  {title}
  {\enquote {\bibinfo {title} {{Negative Casimir entropies in nanoparticle
  interactions}},}\ }\href@noop {} {\bibfield  {journal} {\bibinfo  {journal}
  {J. Condens. Matter Phys.}\ }\textbf {\bibinfo {volume} {27}},\ \bibinfo
  {pages} {214003} (\bibinfo {year} {2015}{\natexlab{b}})}\BibitemShut
  {NoStop}%
\bibitem [{\citenamefont {Milton}\ \emph
  {et~al.}(2017{\natexlab{a}})\citenamefont {Milton}, \citenamefont {Li},
  \citenamefont {Kalauni}, \citenamefont {Parashar}, \citenamefont
  {Gu{\'e}rout}, \citenamefont {Ingold}, \citenamefont {Lambrecht},\ and\
  \citenamefont {Reynaud}}]{milton2017negative}%
  \BibitemOpen
  \bibfield  {author} {\bibinfo {author} {\bibfnamefont {K.~A.}\ \bibnamefont
  {Milton}}, \bibinfo {author} {\bibfnamefont {Y.}~\bibnamefont {Li}}, \bibinfo
  {author} {\bibfnamefont {P.}~\bibnamefont {Kalauni}}, \bibinfo {author}
  {\bibfnamefont {P.}~\bibnamefont {Parashar}}, \bibinfo {author}
  {\bibfnamefont {R.}~\bibnamefont {Gu{\'e}rout}}, \bibinfo {author}
  {\bibfnamefont {G.}~\bibnamefont {Ingold}}, \bibinfo {author} {\bibfnamefont
  {A.}~\bibnamefont {Lambrecht}}, \ and\ \bibinfo {author} {\bibfnamefont
  {S.}~\bibnamefont {Reynaud}},\ }\bibfield  {title} {\enquote {\bibinfo
  {title} {Negative entropies in {Casimir} and {Casimir}-{Polder}
  interactions},}\ }\href@noop {} {\bibfield  {journal} {\bibinfo  {journal}
  {Fortschritte der Phys.}\ }\textbf {\bibinfo {volume} {65}},\ \bibinfo
  {pages} {1600047} (\bibinfo {year} {2017}{\natexlab{a}})}\BibitemShut
  {NoStop}%
\bibitem [{\citenamefont {Li}\ \emph {et~al.}(2016)\citenamefont {Li},
  \citenamefont {Milton}, \citenamefont {Kalauni},\ and\ \citenamefont
  {Parashar}}]{li2016casimir}%
  \BibitemOpen
  \bibfield  {author} {\bibinfo {author} {\bibfnamefont {Y.}~\bibnamefont
  {Li}}, \bibinfo {author} {\bibfnamefont {K.~A.}\ \bibnamefont {Milton}},
  \bibinfo {author} {\bibfnamefont {P.}~\bibnamefont {Kalauni}}, \ and\
  \bibinfo {author} {\bibfnamefont {P.}~\bibnamefont {Parashar}},\ }\bibfield
  {title} {\enquote {\bibinfo {title} {{Casimir self-entropy of an
  electromagnetic thin sheet}},}\ }\href@noop {} {\bibfield  {journal}
  {\bibinfo  {journal} {Phys. Rev. D}\ }\textbf {\bibinfo {volume} {94}},\
  \bibinfo {pages} {085010} (\bibinfo {year} {2016})}\BibitemShut {NoStop}%
\bibitem [{\citenamefont {Milton}\ \emph
  {et~al.}(2017{\natexlab{b}})\citenamefont {Milton}, \citenamefont {Kalauni},
  \citenamefont {Parashar},\ and\ \citenamefont {Li}}]{milton2017casimir}%
  \BibitemOpen
  \bibfield  {author} {\bibinfo {author} {\bibfnamefont {K.~A.}\ \bibnamefont
  {Milton}}, \bibinfo {author} {\bibfnamefont {P.}~\bibnamefont {Kalauni}},
  \bibinfo {author} {\bibfnamefont {P.}~\bibnamefont {Parashar}}, \ and\
  \bibinfo {author} {\bibfnamefont {Y.}~\bibnamefont {Li}},\ }\bibfield
  {title} {\enquote {\bibinfo {title} {{Casimir self-entropy of a spherical
  electromagnetic $\delta$-function shell}},}\ }\href@noop {} {\bibfield
  {journal} {\bibinfo  {journal} {Phys. Rev. D}\ }\textbf {\bibinfo {volume}
  {96}},\ \bibinfo {pages} {085007} (\bibinfo {year}
  {2017}{\natexlab{b}})}\BibitemShut {NoStop}%
\bibitem [{\citenamefont {Bordag}(2018)}]{bordag2018free}%
  \BibitemOpen
  \bibfield  {author} {\bibinfo {author} {\bibfnamefont {M.}~\bibnamefont
  {Bordag}},\ }\bibfield  {title} {\enquote {\bibinfo {title} {Free energy and
  entropy for thin sheets},}\ }\href@noop {} {\bibfield  {journal} {\bibinfo
  {journal} {Phys. Rev. D}\ }\textbf {\bibinfo {volume} {98}},\ \bibinfo
  {pages} {085010} (\bibinfo {year} {2018})}\BibitemShut {NoStop}%
\bibitem [{\citenamefont {Bordag}\ and\ \citenamefont
  {Kirsten}(2018)}]{bordag2018entropy}%
  \BibitemOpen
  \bibfield  {author} {\bibinfo {author} {\bibfnamefont {M.}~\bibnamefont
  {Bordag}}\ and\ \bibinfo {author} {\bibfnamefont {K.}~\bibnamefont
  {Kirsten}},\ }\bibfield  {title} {\enquote {\bibinfo {title} {On the entropy
  of a spherical plasma shell},}\ }\href@noop {} {\bibfield  {journal}
  {\bibinfo  {journal} {J. Phys. A Math. Theor.}\ }\textbf {\bibinfo {volume}
  {51}},\ \bibinfo {pages} {455001} (\bibinfo {year} {2018})}\BibitemShut
  {NoStop}%
\bibitem [{\citenamefont {Milton}\ \emph {et~al.}(2013)\citenamefont {Milton},
  \citenamefont {Abalo}, \citenamefont {Parashar},\ and\ \citenamefont
  {Shajesh}}]{milton2013three}%
  \BibitemOpen
  \bibfield  {author} {\bibinfo {author} {\bibfnamefont {K.~A.}\ \bibnamefont
  {Milton}}, \bibinfo {author} {\bibfnamefont {E.~K.}\ \bibnamefont {Abalo}},
  \bibinfo {author} {\bibfnamefont {P.}~\bibnamefont {Parashar}}, \ and\
  \bibinfo {author} {\bibfnamefont {K.~V.}\ \bibnamefont {Shajesh}},\
  }\bibfield  {title} {\enquote {\bibinfo {title} {{Three-body Casimir-Polder
  interactions}},}\ }\href@noop {} {\bibfield  {journal} {\bibinfo  {journal}
  {Nuovo Cimento. C (Print)}\ }\textbf {\bibinfo {volume} {36}},\ \bibinfo
  {pages} {183--192} (\bibinfo {year} {2013})}\BibitemShut {NoStop}%
\bibitem [{\citenamefont {Milton}\ \emph
  {et~al.}(2015{\natexlab{c}})\citenamefont {Milton}, \citenamefont {Abalo},
  \citenamefont {Parashar}, \citenamefont {Pourtolami}, \citenamefont {Brevik},
  \citenamefont {Ellingsen}, \citenamefont {Buhmann},\ and\ \citenamefont
  {Scheel}}]{milton2015three}%
  \BibitemOpen
  \bibfield  {author} {\bibinfo {author} {\bibfnamefont {K.~A.}\ \bibnamefont
  {Milton}}, \bibinfo {author} {\bibfnamefont {E.~K.}\ \bibnamefont {Abalo}},
  \bibinfo {author} {\bibfnamefont {P.}~\bibnamefont {Parashar}}, \bibinfo
  {author} {\bibfnamefont {N.}~\bibnamefont {Pourtolami}}, \bibinfo {author}
  {\bibfnamefont {I.}~\bibnamefont {Brevik}}, \bibinfo {author} {\bibfnamefont
  {S.~{\AA}}\ \bibnamefont {Ellingsen}}, \bibinfo {author} {\bibfnamefont
  {S.~Y.}\ \bibnamefont {Buhmann}}, \ and\ \bibinfo {author} {\bibfnamefont
  {S.}~\bibnamefont {Scheel}},\ }\bibfield  {title} {\enquote {\bibinfo {title}
  {{Three-body effects in Casimir-Polder repulsion}},}\ }\href@noop {}
  {\bibfield  {journal} {\bibinfo  {journal} {Phys. Rev. A}\ }\textbf {\bibinfo
  {volume} {91}},\ \bibinfo {pages} {042510} (\bibinfo {year}
  {2015}{\natexlab{c}})}\BibitemShut {NoStop}%
\bibitem [{\citenamefont {Li}\ \emph {et~al.}(2022)\citenamefont {Li},
  \citenamefont {Milton}, \citenamefont {Parashar}, \citenamefont {Kennedy},
  \citenamefont {Pourtolami},\ and\ \citenamefont {Guo}}]{li2022casimir}%
  \BibitemOpen
  \bibfield  {author} {\bibinfo {author} {\bibfnamefont {Y.}~\bibnamefont
  {Li}}, \bibinfo {author} {\bibfnamefont {K.~A.}\ \bibnamefont {Milton}},
  \bibinfo {author} {\bibfnamefont {P.}~\bibnamefont {Parashar}}, \bibinfo
  {author} {\bibfnamefont {G.}~\bibnamefont {Kennedy}}, \bibinfo {author}
  {\bibfnamefont {N.}~\bibnamefont {Pourtolami}}, \ and\ \bibinfo {author}
  {\bibfnamefont {X.}~\bibnamefont {Guo}},\ }\bibfield  {title} {\enquote
  {\bibinfo {title} {{Casimir self-entropy of nanoparticles with classical
  polarizabilities: Electromagnetic Field Fluctuations}},}\ }\href@noop {}
  {\bibfield  {journal} {\bibinfo  {journal} {Phys. Rev. D}\ }\textbf {\bibinfo
  {volume} {106}},\ \bibinfo {pages} {036002} (\bibinfo {year}
  {2022})}\BibitemShut {NoStop}%
\bibitem [{\citenamefont {Li}\ \emph {et~al.}(2021)\citenamefont {Li},
  \citenamefont {Milton}, \citenamefont {Parashar},\ and\ \citenamefont
  {Hong}}]{li2021negativity}%
  \BibitemOpen
  \bibfield  {author} {\bibinfo {author} {\bibfnamefont {Y.}~\bibnamefont
  {Li}}, \bibinfo {author} {\bibfnamefont {K.~A.}\ \bibnamefont {Milton}},
  \bibinfo {author} {\bibfnamefont {P}~\bibnamefont {Parashar}}, \ and\
  \bibinfo {author} {\bibfnamefont {L}~\bibnamefont {Hong}},\ }\bibfield
  {title} {\enquote {\bibinfo {title} {{Negativity of the Casimir self-entropy
  in spherical geometries}},}\ }\href@noop {} {\bibfield  {journal} {\bibinfo
  {journal} {Entropy}\ }\textbf {\bibinfo {volume} {23}},\ \bibinfo {pages}
  {214} (\bibinfo {year} {2021})}\BibitemShut {NoStop}%
\bibitem [{\citenamefont {Milton}\ \emph {et~al.}(2019)\citenamefont {Milton},
  \citenamefont {Kalauni}, \citenamefont {Parashar},\ and\ \citenamefont
  {Y.}}]{milton2017remarks}%
  \BibitemOpen
  \bibfield  {author} {\bibinfo {author} {\bibfnamefont {K.~A.}\ \bibnamefont
  {Milton}}, \bibinfo {author} {\bibfnamefont {P.}~\bibnamefont {Kalauni}},
  \bibinfo {author} {\bibfnamefont {P}~\bibnamefont {Parashar}}, \ and\
  \bibinfo {author} {\bibfnamefont {Li}~\bibnamefont {Y.}},\ }\bibfield
  {title} {\enquote {\bibinfo {title} {{Remarks on the Casimir self-entropy of
  a spherical electromagnetic $delta$-function shell}},}\ }\href@noop {}
  {\bibfield  {journal} {\bibinfo  {journal} {Phys. Rev. D}\ }\textbf {\bibinfo
  {volume} {99}},\ \bibinfo {pages} {045013} (\bibinfo {year}
  {2019})}\BibitemShut {NoStop}%
\bibitem [{\citenamefont {Guo}\ \emph {et~al.}(2021)\citenamefont {Guo},
  \citenamefont {Milton}, \citenamefont {Kennedy}, \citenamefont {McNulty},
  \citenamefont {Pourtolami},\ and\ \citenamefont {Li}}]{guo2021energetics}%
  \BibitemOpen
  \bibfield  {author} {\bibinfo {author} {\bibfnamefont {X.}~\bibnamefont
  {Guo}}, \bibinfo {author} {\bibfnamefont {K.~A.}\ \bibnamefont {Milton}},
  \bibinfo {author} {\bibfnamefont {G.}~\bibnamefont {Kennedy}}, \bibinfo
  {author} {\bibfnamefont {W.~P.}\ \bibnamefont {McNulty}}, \bibinfo {author}
  {\bibfnamefont {N.}~\bibnamefont {Pourtolami}}, \ and\ \bibinfo {author}
  {\bibfnamefont {Y.}~\bibnamefont {Li}},\ }\bibfield  {title} {\enquote
  {\bibinfo {title} {{Energetics of quantum vacuum friction: Field
  fluctuations}},}\ }\href@noop {} {\bibfield  {journal} {\bibinfo  {journal}
  {Phys. Rev. D}\ }\textbf {\bibinfo {volume} {104}},\ \bibinfo {pages}
  {116006} (\bibinfo {year} {2021})}\BibitemShut {NoStop}%
\bibitem [{\citenamefont {Axilrod}\ and\ \citenamefont
  {Teller}(1943)}]{Axilrod}%
  \BibitemOpen
  \bibfield  {author} {\bibinfo {author} {\bibfnamefont {B.~M.}\ \bibnamefont
  {Axilrod}}\ and\ \bibinfo {author} {\bibfnamefont {E}~\bibnamefont
  {Teller}},\ }\bibfield  {title} {\enquote {\bibinfo {title} {{Interaction of
  the van der Waals type between three atoms}},}\ }\href@noop {} {\bibfield
  {journal} {\bibinfo  {journal} {J. Chem. Phys.}\ }\textbf {\bibinfo {volume}
  {11}},\ \bibinfo {pages} {299--300} (\bibinfo {year} {1943})}\BibitemShut
  {NoStop}%
\bibitem [{\citenamefont {Milton}\ \emph {et~al.}(2020)\citenamefont {Milton},
  \citenamefont {Parashar}, \citenamefont {Brevik},\ and\ \citenamefont
  {Kennedy}}]{milton2020self}%
  \BibitemOpen
  \bibfield  {author} {\bibinfo {author} {\bibfnamefont {K.~A.}\ \bibnamefont
  {Milton}}, \bibinfo {author} {\bibfnamefont {P.}~\bibnamefont {Parashar}},
  \bibinfo {author} {\bibfnamefont {I.}~\bibnamefont {Brevik}}, \ and\ \bibinfo
  {author} {\bibfnamefont {G.}~\bibnamefont {Kennedy}},\ }\bibfield  {title}
  {\enquote {\bibinfo {title} {{Self-stress on a dielectric ball and
  Casimir-Polder forces}},}\ }\href@noop {} {\bibfield  {journal} {\bibinfo
  {journal} {Annals of Physics}\ }\textbf {\bibinfo {volume} {412}},\ \bibinfo
  {pages} {168008} (\bibinfo {year} {2020})}\BibitemShut {NoStop}%
\bibitem [{\citenamefont {Cao}\ \emph {et~al.}(2019)\citenamefont {Cao},
  \citenamefont {Chern}, \citenamefont {Dennis},\ and\ \citenamefont
  {Brown}}]{cao2019measuring}%
  \BibitemOpen
  \bibfield  {author} {\bibinfo {author} {\bibfnamefont {W.}~\bibnamefont
  {Cao}}, \bibinfo {author} {\bibfnamefont {M.}~\bibnamefont {Chern}}, \bibinfo
  {author} {\bibfnamefont {A.~M.}\ \bibnamefont {Dennis}}, \ and\ \bibinfo
  {author} {\bibfnamefont {K.~A.}\ \bibnamefont {Brown}},\ }\bibfield  {title}
  {\enquote {\bibinfo {title} {Measuring nanoparticle polarizability using
  fluorescence microscopy},}\ }\href@noop {} {\bibfield  {journal} {\bibinfo
  {journal} {Nano Lett.}\ }\textbf {\bibinfo {volume} {19}},\ \bibinfo {pages}
  {5762--5768} (\bibinfo {year} {2019})}\BibitemShut {NoStop}%
\bibitem [{\citenamefont {Mader}\ \emph {et~al.}(2022)\citenamefont {Mader},
  \citenamefont {Benedikter}, \citenamefont {Husel}, \citenamefont
  {H\"{a}nsch},\ and\ \citenamefont {Hunger}}]{mader2022quantitative}%
  \BibitemOpen
  \bibfield  {author} {\bibinfo {author} {\bibfnamefont {M.}~\bibnamefont
  {Mader}}, \bibinfo {author} {\bibfnamefont {J.}~\bibnamefont {Benedikter}},
  \bibinfo {author} {\bibfnamefont {L.}~\bibnamefont {Husel}}, \bibinfo
  {author} {\bibfnamefont {T.~W.}\ \bibnamefont {H\"{a}nsch}}, \ and\ \bibinfo
  {author} {\bibfnamefont {D.}~\bibnamefont {Hunger}},\ }\bibfield  {title}
  {\enquote {\bibinfo {title} {{Quantitative Determination of the Complex
  Polarizability of Individual Nanoparticles by Scanning Cavity Microscopy}},}\
  }\href@noop {} {\bibfield  {journal} {\bibinfo  {journal} {ACS Photonics}\
  }\textbf {\bibinfo {volume} {9}},\ \bibinfo {pages} {466--473} (\bibinfo
  {year} {2022})}\BibitemShut {NoStop}%
\bibitem [{\citenamefont {Bordag}\ and\ \citenamefont
  {Pirozhenko}(2017)}]{bordag2017casimir}%
  \BibitemOpen
  \bibfield  {author} {\bibinfo {author} {\bibfnamefont {M.}~\bibnamefont
  {Bordag}}\ and\ \bibinfo {author} {\bibfnamefont {I.~G.}\ \bibnamefont
  {Pirozhenko}},\ }\bibfield  {title} {\enquote {\bibinfo {title} {{Casimir
  effect for Dirac lattices}},}\ }\href@noop {} {\bibfield  {journal} {\bibinfo
   {journal} {Phys. Rev. D}\ }\textbf {\bibinfo {volume} {95}},\ \bibinfo
  {pages} {056017} (\bibinfo {year} {2017})}\BibitemShut {NoStop}%
\bibitem [{\citenamefont {Pirozhenko}(2020)}]{pirozhenko2020finite}%
  \BibitemOpen
  \bibfield  {author} {\bibinfo {author} {\bibfnamefont {I.}~\bibnamefont
  {Pirozhenko}},\ }\bibfield  {title} {\enquote {\bibinfo {title} {{On finite
  temperature Casimir effect for Dirac lattices}},}\ }\href@noop {} {\bibfield
  {journal} {\bibinfo  {journal} {Modern Physics Letters A}\ }\textbf {\bibinfo
  {volume} {35}},\ \bibinfo {pages} {2040019} (\bibinfo {year}
  {2020})}\BibitemShut {NoStop}%
\bibitem [{\citenamefont {Sircar}\ \emph {et~al.}(2020)\citenamefont {Sircar},
  \citenamefont {Patra},\ and\ \citenamefont {Batra}}]{sircar2020casimir}%
  \BibitemOpen
  \bibfield  {author} {\bibinfo {author} {\bibfnamefont {A.}~\bibnamefont
  {Sircar}}, \bibinfo {author} {\bibfnamefont {P.~K.}\ \bibnamefont {Patra}}, \
  and\ \bibinfo {author} {\bibfnamefont {R.~C.}\ \bibnamefont {Batra}},\
  }\bibfield  {title} {\enquote {\bibinfo {title} {Casimir force and its
  effects on pull-in instability modelled using molecular dynamics
  simulations},}\ }\href@noop {} {\bibfield  {journal} {\bibinfo  {journal}
  {Proc. Math. Phys. Eng. Sci.}\ }\textbf {\bibinfo {volume} {476}},\ \bibinfo
  {pages} {20200311} (\bibinfo {year} {2020})}\BibitemShut {NoStop}%
\end{thebibliography}%

\end{document}